%% file: main_mnras.tex
\documentclass[fleqn,usenatbib]{mnras}

\usepackage{newtxtext,newtxmath}

\usepackage[T1]{fontenc}

\DeclareRobustCommand{\VAN}[3]{#2}
\let\VANthebibliography\thebibliography
\def\thebibliography{\DeclareRobustCommand{\VAN}[3]{##3}\VANthebibliography}

\usepackage{graphicx}	
\usepackage{amsmath}	
\usepackage{comment}
\usepackage{xspace} 
\usepackage{mathtools}
\usepackage{hyperref}
\usepackage{float}
\usepackage{bm}
\usepackage{pdflscape}
\usepackage{pdfpages}
\usepackage{subcaption}
\usepackage{glossaries}
\DeclareFontFamily{U}{mathx}{}
\DeclareFontShape{U}{mathx}{m}{n}{<-> mathx10}{}
\DeclareSymbolFont{mathx}{U}{mathx}{m}{n}
\DeclareMathAccent{\widecheck}{0}{mathx}{"71}

\usepackage{multirow} \usepackage{threeparttable}
\usepackage{booktabs} \usepackage{stfloats}

\newcommand{\thisalgo}{{\texttt{MODEL\&CO}}\xspace} 

\newcommand*{\V}[1]{\boldsymbol{#1}}   
\newcommand*{\M}[1]{\mathbf{#1}}       
\newcommand*{\TransposeLetter}{\hspace*{-.25ex}\top\hspace*{-.25ex}}
\newcommand*{\T}{^{\TransposeLetter}} 
\newcommand*{\eqd}{\,{\buildrel d \over =}\,}
\DeclareFontFamily{U}{mathx}{\hyphenchar\font45}
\DeclareFontShape{U}{mathx}{m}{n}{<-> mathx10}{}
\DeclareSymbolFont{mathx}{U}{mathx}{m}{n}

\DeclarePairedDelimiterX{\Paren}[1]{(}{)}{#1}
\DeclarePairedDelimiterX{\Brace}[1]{\{}{\}}{#1}
\DeclarePairedDelimiterX{\Brack}[1]{[}{]}{#1}
\DeclarePairedDelimiterX{\Abs}[1]{\rvert}{\lvert}{#1}
\DeclarePairedDelimiterX{\Norm}[1]{\lVert}{\rVert}{#1}
\DeclarePairedDelimiterX{\Avg}[1]{\langle}{\rangle}{#1}
\DeclarePairedDelimiterX{\Round}[1]{\lfloor}{\rceil}{#1}
\DeclarePairedDelimiterX{\Floor}[1]{\lfloor}{\rfloor}{#1}
\DeclarePairedDelimiterX{\Ceil}[1]{\lceil}{\rceil}{#1}
\DeclarePairedDelimiterX{\Inner}[2]{\langle}{\rangle}{#1,#2}

\DeclareMathOperator*{\argmin}{arg\,min}

\DeclareMathOperator{\tr}{tr}

\DeclarePairedDelimiterXPP{\Expect}[1]{\mathbb{E}}(){}{#1}

\usepackage[squaren,Gray]{SIunits}

\usepackage{numprint}

\makeatletter
\def\widebreve{\mathpalette\wide@breve}
\def\wide@breve#1#2{\sbox\z@{$#1#2$}%
     \mathop{\vbox{\m@th\ialign{##\crcr
\kern0.08em\brevefill#1{0.8\wd\z@}\crcr\noalign{\nointerlineskip}%
                    $\hss#1#2\hss$\crcr}}}\limits}
\def\brevefill#1#2{$\m@th\sbox\tw@{$#1($}%
  \hss\resizebox{#2}{\wd\tw@}{\rotatebox[origin=c]{90}{\upshape(}}\hss$}
\makeatletter

\title[Multi-Observations DEep Learning \& COvariance]{MODEL\&CO: Exoplanet detection in angular differential imaging \\by learning across multiple observations}

\author[T. Bodrito et al.]{
Théo Bodrito$^{1}$\thanks{E-mail: theo.bodrito@inria.fr},  Olivier Flasseur$^{2}$, Julien Mairal$^{3}$, Jean Ponce$^{1,4}$, \newauthor Maud Langlois$^{2}$, Anne-Marie Lagrange$^{5,6}$
\\
$^{1}$Département d'Informatique de l'{\'E}cole Normale Supérieure (ENS-PSL, CNRS, Inria), France\\
$^{2}$Centre de Recherche Astrophysique de Lyon, CNRS, Univ. de Lyon, Univ. Claude Bernard Lyon 1, ENS de Lyon, France\\
$^{3}$Univ. Grenoble Alpes, Inria, CNRS, Grenoble INP, LJK, 38000 Grenoble, France\\
$^{4}$Courant Institute of Mathematical Sciences, Center for Data Science, New York Univ., USA\\
$^{5}$Laboratoire d'{\'E}tudes Spatiales et d'Instrumentation en Astrophysique, Observatoire de Paris, Univ. PSL, Sorbonne Univ., Univ. Paris Diderot, France\\
$^{6}$Univ. Grenoble Alpes, Institut de Planétologie et d'Astrophysique de Grenoble, France
}

\pubyear{2024}

\input{./glossary.tex}

\begin{document}
\label{firstpage}
\pagerange{\pageref{firstpage}--\pageref{lastpage}}
\maketitle

\begin{abstract}
Direct imaging of exoplanets is particularly challenging due to the high contrast between the planet and the star luminosities, and their small angular separation.
In addition to tailored instrumental facilities implementing adaptive optics and coronagraphy, post-processing methods combining several images recorded in pupil tracking mode are needed to attenuate the nuisances corrupting the signals of interest.
Most of these post-processing methods build a model of the nuisances from the target observations themselves, resulting in strongly limited detection sensitivity at short angular separations due to the lack of angular diversity.
To address this issue, we propose to build the nuisance model from an archive of multiple observations by leveraging supervised deep learning techniques.
The proposed approach casts the detection problem as a reconstruction task and captures the structure of the nuisance from two complementary representations of the data.
Unlike methods inspired by reference differential imaging, the proposed model is highly non-linear and does not resort to explicit image-to-image similarity measurements and subtractions.
The proposed approach also encompasses statistical modeling of learnable spatial features.
The latter is beneficial to improve both the detection sensitivity and the robustness against heterogeneous data.
We apply the proposed algorithm to several datasets from the VLT/SPHERE instrument, and demonstrate a superior precision-recall trade-off compared to the PACO algorithm.
Interestingly, the gain is especially important when the diversity induced by ADI is the most limited, thus supporting the ability of the proposed approach to learn information across multiple observations.
\end{abstract}


\maketitle 

\input{introduction}

\input{method}

\input{results}

\input{conclusion}

\section*{Acknowledgements}

\noindent We thank the Referee for their careful reading of the manuscript as well as their insightful comments and suggestions.
We also thank Julien Milli (IPAG, Grenoble, France) for the fruitful discussions regarding the monitoring of the observing conditions at Paranal Observatory. 

This work was funded by the French government under management of National Research Agency (ANR) as part of the \textit{Investissements d'Avenir} program with grant agreements ANR19-P3IA0001 (PRAIRIE 3IA Institute) and ANR-19-P3IA-0003 (MIAI 3IA Institute), by the European Research Council (ERC) under grant agreements 101087696 (APHELEIA project), the Inria/NYU collaboration, and the Louis Vuitton/ENS chair on artificial intelligence.
This project was supported by the European Research Council (ERC) under the European Union's Horizon 2020 research and innovation programme (COBREX; grant agreement n° 885593) "

This work was granted access to the HPC resources of IDRIS under the allocation 2022-AD011013643 made by GENCI.
 
TB, OF, JM, and JP conceived and designed the method as well as the analysis presented in this paper. TB developed, tested, and implemented the algorithm. ML and AML provided the data archive. TB and OF selected and annotated the database. TB and OF performed the analysis of the results. TB, OF, JM, JP, ML, and AML wrote the manuscript.

\section*{Data Availability}
 
The raw data used in this article are freely available on the ESO archive facility at \href{http://archive.eso.org/eso/eso\_archive\_main.html}{http://archive.eso.org/eso/eso\_archive\_main.html}. They were pre-reduced with the SPHERE Data Centre, jointly operated by OSUG/IPAG (Grenoble), PYTHEAS/LAM/CESAM (Marseille), OCA/Lagrange (Nice), Observatoire de Paris/LESIA (Paris), and Observatoire de Lyon/CRAL (Lyon, France). 


\bibliographystyle{aa} 
\bibliography{biblio.bib} 

\appendix
\input{appendix.tex}




\end{document}

%% file: glossary.tex
\makeglossaries
\newglossaryentry{empirical_risk_minimization}
{
    name=Empirical risk minimization,
    text=empirical risk minimization,
    description={
    We consider a family of prediction functions $f_\theta: \mathcal{Y} \to \mathcal{X}$ parametrized by $\theta$ in $\Theta$, and denoted by $\ell: \mathcal{X}^2 \to \mathbb{R}$ the so-called \textit{loss function}
      measuring the difference between the prediction $f_\theta(\V y)$ and the true value $\V x$.
      Let $\mathcal{Z}$ represent the set of all possible input-output pairs $(\V y, \V x)$.
      The \textit{true risk}, defined as $\mathcal{R}(f_\theta) = \mathbb{E}_\mathcal{Z}[\ell(f_\theta(\V y), \V x)]$ is the quantity we aim to minimize.
      However, since $\mathcal{Z}$ is unknown in practice, we instead work with a finite subset $\mathcal{D}$ of $\mathcal{Z}$ containing $n$ samples.
      For this dataset, one typically computes the \textit{empirical risk} $\mathcal{\hat{R}}(f_\theta) = \frac{1}{n}\sum_{(y, x) \in \mathcal{D}} \ell(f(\V y), \V x)$ which approximates the true risk.
      \textit{Empirical risk minimization} aims to find the parameters $\hat{\theta} = \argmin_{\theta \in \Theta} \mathcal{\hat{R}}(f_\theta)$,
      and is typically achieved via \gls{stochastic_gradient_descent}}
}

\newglossaryentry{stochastic_gradient_descent}
{
    name=Stochastic gradient descent,
    text=stochastic gradient descent,
    description={
      Consider a family of parameters $\Theta$, and an objective function with the form $\mathcal{L}(\theta) = \frac{1}{n} \sum_i \mathcal{L}_i(\theta)$ to be minimized.
      Consider a family of parameters $\Theta$, and an objective function $\mathcal{L}: \Theta \to \mathbb{R}$ with the form $\mathcal{L}(\theta) = \frac{1}{n} \sum_i \mathcal{L}_i(\theta)$ to be minimized.
      In \gls{empirical_risk_minimization}, each term $\mathcal{L}_i$ corresponds to the $i$-th sample in the dataset of size $n$.
      \textit{Gradient descent} is an iterative method to minimize an objective function, an its iterations write 
      $\theta_{k+1} = \theta_k - \eta \nabla \mathcal{L}(\theta_k)$, where $\nabla \mathcal{L} (\theta) = \frac{1}{n} \sum_i \nabla \mathcal{L}_i(\theta)$ represents the gradient of the objective function in $\theta$, and $\eta$ is a tunable parameter known as the \textit{learning rate}.
      In \textit{stochastic gradient descent}, the true gradient is approximated by the gradient of a single sample $i_k$ randomly chosen at each iteration $k$: $\theta_{k+1} = \theta_k - \eta \nabla \mathcal{L}_{i_k}(\theta_k)$.
      A compromise between computing the true gradient and the gradient at a single sample is to compute the mean gradient over a small set of samples (called a \textit{mini-batch}) at each step.
      Stochastic gradient descent and its variants are widely used for optimizing the parameters of \gls{deep_learning} models}
}

\newglossaryentry{deep_learning}
{
    name=Deep learning,
    text=deep learning,
    description={
      refers to a special class of parametrized functions, structured in multiple \textit{layers}.
    The most basic layer is the \textit{fully connected layer}, which performs a non-linear transformation of the previous layer's output, denoted by $y^{(k)} \in \mathbb{R}^d$. 
    The forward pass through the $k$-the layer is defined by:
      \begin{align}
        x^{(k+1)} &= W^{(k)} y^{(k)} + b^{(k)}\\
        y^{(k+1)} &= \sigma(x^{(k+1)})
      \end{align}
      where $W^{(k)} \in \mathbb{R}^{d \times d}$ and $b^{(k)} \in \mathbb{R}^{d}$ are learnable parameters associated with layer $k$, and $\sigma$ denotes a non-linear function.
      The overall function, or \textit{neural network}, is defined as $f_\theta(y^{(0)}) = x^{(K)}$, where $K$ is the number of layers, and $\theta = \{W^{(0)}, b^{(0)}, \cdots, W^{(K-1)}, b^{(K-1)}\}$ is the set of all learnable parameters}
}

\newglossaryentry{residual}
{
    name=Residual layer,
    text=residual,
    description={
      Alternative \gls{deep_learning} architecture, where each layer $k$ is defined as:
      \begin{align}
        x^{(k+1)} &= W^{(k)} y^{(k)} + b^{(k)}\\
        y^{(k+1)} &= y^{(k)} + \sigma(x^{(k+1)})
      \end{align}
    This architecture has been shown to improve and stabilize the training of deep neural networks}
}

\newglossaryentry{cnn}
{
    name=Convolutional neural network,
    text=convolutional neural network,
    description={
      A convolutional neural network (CNN) is a type of \gls{deep_learning} model specifically designed for processing and analyzing grid-like data, such as images. 
      A convolutional layer applies local filters to the input, capturing patterns such as edges and textures. 
      CNNs typically consist of multiple convolutional layers, often coupled with fully connected layers. 
      The ability to learn and generalize spatial patterns makes CNNs highly effective for image classification, object detection, and other visual recognition tasks}
}

\newglossaryentry{hadamard_product}
{
    name=Hadamard product,
    description={
      Element-wise multiplication}
}

\newglossaryentry{iid}
{
    name=I.i.d.,
    text=i.i.d.,
    description={
  independent identically distributed}
}

\newglossaryentry{cholesky_factorization}
{
    name=Cholesky factorization,
    description={
      Decomposition of a real symmetric positive-definite matrix $M = LL^\top$ where $L$ is a lower triangular matrix with positive diagonal entries}
}

\newglossaryentry{normalization_layer}
{
    name=Normalization layer,
    text=normalization layer,
    description={
A layer in deep learning models that standardizes the inputs to a specific scale, typically by adjusting and scaling the features to have a mean of zero and a standard deviation of one. 
This normalization process helps stabilize and accelerate training, and enables higher learning rates. 
Examples include batch, layer, instance and group normalization, each applied in different contexts such as across mini-batches, within individual layers, or per instance.
\textit{Batch normalization}: adjusts and scales each feature independently within a mini-batch to have a mean of zero and a standard deviation of one. 
\textit{Layer normalization}: normalizes across the features of a single data point instead of across a mini-batch}
}

\newglossaryentry{shrinkage}
{
    name=Shrinkage (covariance),
    text=shrinkage,
    description={
A regularization technique applied to the sample covariance matrix, used when estimating covariance from limited data. 
Shrinkage combines the sample covariance matrix with a simpler matrix (such as the diagonal covariance matrix) to create a more robust estimator that mitigates overfitting and improves generalization}
}

\newglossaryentry{convex_combination}
{
    name=Convex combination,
    text=convex combination,
    description={
A linear combination of points where all coefficients are non-negative and sum to one. In mathematical terms, for points $x_1, x_2, \dots, x_n$, a convex combination is $y = \alpha_1 x_1 + \alpha_2 x_2 + \dots + \alpha_n x_n$, where $\alpha_i \geq 0$ and $\sum_i \alpha_i = 1$}
}

\newglossaryentry{bias_variance_tradeoff}
{
    name=Bias-variance trade-off,
    text=bias-variance trade-off,
    description={
      A key concept in machine learning and statistics that describes the trade-off between \textit{bias} (error due to overly simplistic models) and \textit{variance} (error due to overly complex models). 
High bias can lead to underfitting, while high variance can lead to overfitting. Achieving a balance between the two is crucial for model generalization}
}

\newglossaryentry{unet}
{
    name=U-Net,
    description={
A type of convolutional neural network architecture designed for image segmentation tasks. 
It has an encoder-decoder structure with skip connections between corresponding layers of the encoder and decoder, allowing it to capture both global context and fine details in images. 
It is widely used in medical imaging and other applications requiring precise localization}
}

\newglossaryentry{ablation_study}
{
    name=Ablation study,
    text=ablation study,
    description={
A research method in the machine learning community used to analyze the impact of different components or features of a model. 
Ablation involves systematically removing or disabling parts of a model (such as specific layers or parameters) to understand their contribution to the overall performance and identify the most important components}
}

\newglossaryentry{trace}
{
    name=Trace,
    text=trace,
    description={
The sum of the diagonal elements of a square matrix}
}

\newglossaryentry{rectified_linear_unit}
{
    name=Rectified Linear Unit (ReLU),
    description={
A widely-used activation function in neural networks that is defined as $\text{ReLU}(x) = \max(0, x)$. 
ReLU introduces non-linearity to the model while being computationally efficient, as it simply sets negative inputs to zero. 
ReLU helps mitigate the vanishing gradient problem in deep networks, making it a popular choice in modern deep learning architectures}
}

\newglossaryentry{probability_integral_transform_theorem}
{
    name=Probability integral transform theorem,
    text=probability integral transform theorem,
    description={
    states that if a random variable $X$ has a continuous cumulative distribution function (CDF) $F_X(x)$, then the random variable $U = F_X(X)$ is uniformly distributed on the interval $[0, 1]$. 
    This theorem is important in statistical methods, where uniformly distributed random variables can be transformed into variables that follow a specified distribution using the inverse of the CDF}
}

\newglossaryentry{equality_distribution}
{
    name=Equality in distribution,
    text=equality in distribution,
    description={
  Two random variables $X$ and $Y$ are equal in distribution if they have the same probability distribution.
  This is commonly denoted by $X \eqd Y$}
}

%% file: introduction.tex
\section{Introduction} 
\label{sec:introduction}

\noindent Direct imaging is an observational technique \citep{roddier1999adaptive, traub2010direct,bowler2016imaging} notably used for scrutinizing the circumstellar environment of nearby stars.
This has thus far led to the detection and spectroscopic characterization of several dozen young, giant, hot, and self-luminous exoplanets in the infrared \citep{chauvin2004giant, chauvin2005giant, marois2008direct, lagrange2009probable}. 
For large-scale surveys, see \cite{nielsen2019gemini, desidera2021sphere, langlois2021sphere}, and for recent reviews, see \cite{pueyo2018direct, macintosh2018gemini, currie2022direct,follette2023introduction}.
The primary challenges of direct imaging lie in the required high angular resolution and high contrast between the host star and the exoplanets, which make detection particularly difficult.
In this context, most leading ground-based observational facilities are equipped with instruments tailored for exoplanet imaging (e.g., Gemini/GPI; \cite{macintosh2014first}, Magellan/MagAO; \cite{morzinski2014magao}, Keck/NIRC2; \cite{castella2016commissioning}, SUBARA/SCExAO; \cite{jovanovic2015subaru}, VLT/SPHERE; \cite{beuzit2019sphere} for the most recent ones). 
They integrate specific optical devices such as coronagraphic masks \citep{macintosh2014first, beuzit2019sphere} and (extreme) adaptive optics \citep{davies2012adaptive,milli2016adaptive} to respectively block out part of the stellar light and correct in real time for the wave-front distortion induced by atmospheric turbulence.
However, the residual optical aberrations of the instrument lead, due to diffraction, to temporally-varying patterns in the sensor plane known as \textit{speckles} \citep{fitzgerald2006speckle}.
Classical sources of noise also add to the speckles to form a strong, spatially correlated, and non-stationary nuisance component.
This nuisance and its random fluctuations are the main bottlenecks for direct imaging. 
As an illustration, the fluctuating residual speckles dominate the signals of the sought objects, with residual contrast around $10^3-10^4$ in the images near the star, even under optimal observing conditions.
To further improve the detection performance, dedicated observation strategies are thus employed to discriminate between the signal of interest and the nuisances. 
Among them, angular differential imaging (ADI; \cite{marois2006angular}) is used routinely in direct imaging: measurements are taken with the field derotator of the telescope tuned to keep the telescope pupil stable while the field of view rotates around the target star.
The rotation of the Earth thus induces an apparent motion of the off-axis sources, while the speckles remain quasi-static. 
This diversity can be leveraged by dedicated post-processing algorithms to unmix the source signals from the nuisance.

Given the high variability among observations, most post-processing algorithms construct a unique model of the nuisance for each individual observation
\citep{pueyo2018direct,cantalloube2020exoplanet}. 
We refer to this type of approach as \textit{observation-dependent}.
Among these methods, the first and perhaps most commonly used class empirically constructs an empirical reference model of the nuisance, which is then subtracted from each image of the target observation. 
Subsequently, the parallactic rotation of the off-axis sources is compensated for, and the residuals are co-added to recover the signal of interest.
The reference model of the nuisance can be estimated using various techniques, such as performing a temporal mean or median (cADI; \cite{marois2006angular,lagrange2009probable}), a (local) linear combination of images (LOCI-based algorithms; \cite{lafreniere2007new,marois2013tloci,marois2014gpi,wahhaj2015improving}) or a (local) principle component analysis (PCA-based algorithms; \cite{soummer2012detection,amara2012pynpoint}). 

Over the years, more sophisticated subtraction-based methods have been developed: LLSG \citep{gonzalez2016low} decomposes an ADI sequence into low-rank, sparse, and Gaussian noise components; 
TRAP \citep{samland2021trap} and HSR \citep{gebhard2022half} perform pixel-wise time series regression with independent features to discriminate off-axis sources from the nuisance; and RSM \citep{dahlqvist2020regime,dahlqvist2021improving} improves upon previous methods by conducting a finer-grained analysis of the residuals. 
Another class of observation-dependent algorithms (e.g., ANDROMEDA \cite{cantalloube2015direct}, FMMF \cite{ruffio2017improving}) adopts a statistical detection framework based on matched filtering. 
Among these, the PACO algorithm \citep{flasseur2018exoplanet,flasseur2020robustness} models collections of temporal patches as mixtures of scaled multivariate Gaussians, deriving a closed-form expression of the posterior distribution accounting for the spatial correlations of the data. 
Given recent advances in machine and \gls{deep_learning}, fully data-driven approaches based on supervised learning have also been explored. 
SODINN \citep{gonzalez2018supervised} performs binary classification on patches using a random forest or a \gls{cnn}.
NA-SODINN \citep{cantero2023sodinn} improves upon SODINN by training multiple models for different noise regimes, typically background-limited or speckle-limited.
The hybrid deep PACO algorithm \citep{flasseur2023combining,flasseur2024deep} performs supervised binary classification on full images, spatially whitened using the statistical parameters inferred by PACO.
 
\medskip

\noindent All observation-dependent approaches suffer from a common limitation: the detection sensitivity stays upper-bounded at short angular separations by the lack of angular diversity. 
and the possible presence of objects of interest.
As an illustration, a gap in contrast by a factor 10 to 30 
remains between the detection sensitivity reached by the deep PACO algorithm and the theoretical ultimate detection sensitivity driven by the photon noise limit \citep{flasseur2024deep}. 
While this gap quickly decreases with the angular separation and is almost null beyond 0.5'', it strongly limits our capability to image exoplanets of a few Jovian masses in the closest stellar vicinity. 
As a comparison, the indirect radial velocity method is much more sensitive in this regime, and has led to the discovery of several Jupiter and Neptune analogues below 10 au, i.e., closer to the water ice line where the bulk of such exoplanets is expected to be \citep{fernandes2019hints,fulton2021california}. 
From a data-processing standpoint, such limitations in detection sensitivity with direct imaging are the results of two effects occurring simultaneously near the host star: 
(i) the nuisance is very strong and displays larger temporal fluctuations than farther away, and 
(ii) the apparent rotation of the off-axis objects induced by ADI is not sufficient, leading to substantial \textit{self-subtraction} where part of the off-axis signal is inadvertently included in the nuisance contribution.

After more than a decade of exploitation of direct imaging instruments, new possibilities have emerged to address these issues. 
In particular, hundreds of observations have been collected from large surveys such as the SpHere INfrared Exoplanets survey (SHINE; \cite{desidera2021sphere}) with SPHERE and the Gemini Planet Imager Exoplanet Survey (GPIES; \cite{nielsen2019gemini}) with GPI. 
The diversity and redundancy of observations within this extensive database represent valuable information that remains largely under-exploited.
Two primary research directions are currently being explored to mitigate the limitations of observation-dependent approaches through dedicated post-processing strategies. 
In that context and still focusing on ADI, two main lines of research are currently investigated to mitigate the limitations of observation-dependent approaches through dedicated post-processing strategies. 
Both approaches leverage multiple direct imaging observations to create an \textit{observation-independent} model, yet they differ in how they exploit this added diversity.
The first category of methods utilizes data fusion techniques to coherently combine the signals of the objects of interest \citep{le2020k, thompson2022deep, dallant2023pacome} from multiple epochs. 
These epochs represent several observations of the same target star recorded at different times, potentially spanning several years.
These methods are essential for determining whether a detected source is gravitationally bound to the host star and for estimating the parameters of its orbit. 
However, they are quite costly for the detection task, as they require multiple observations of the same target stars.
The second category of methods, leveraging multiple direct imaging observations, focuses on more accurately modeling the nuisance component. 
In particular, they aim to mitigate the detrimental self-subtraction phenomenon that occurs at short angular separations.
Such a model can be constructed using observations of reference stars, where potential off-axis sources are expected to not spatially coincide with those of the target star.
This is the general principle of approaches based on reference differential imaging (RDI; \cite{smith1984circumstellar, lafreniere2009hst, ren2018non,xuan2018characterizing,ruane2019reference,bohn2020two,xie2022reference,ren2023karhunen}). 
The required reference observations can be selected from extensive surveys conducted using the same instrument. 
Alternatively, another technique known as star-hopping, involves alternating between observing the target star and a reference star within the same observing sequence \citep{wahhaj2021search}.
This approach reduces the time allocated to observing the target of interest, which can negatively impact detection performances in regions where the speckle nuisance is properly captured.
Given our objective to develop a new approach suitable for deployment in both current and future large-scale observation programs, we emphasize survey databases as a crucial additional source of diversity to complement ADI.
Regardless of how the references are constructed, RDI relies on the stability of the instrument to ensure that residual aberrations evolve slowly, resulting in consistent speckle structures across observations.
Thus, RDI searches for (potentially local) similarities between images of the target observation and the multiple reference observations. 
The nuisance model in existing RDI post-processing methods is typically either a linear combination of the selected images themselves or their low-rank representation (i.e., feature vectors) obtained, for example, through PCA. 
The resulting observation-independent model of the nuisance is subsequently subtracted from the target observation, following a framework similar to classical subtraction-based algorithms.
RDI is used routinely to process space-based observations (see e.g., \cite{choquet2014archival,schneider2014probing,schneider2016deep,hagan2018alice,ren2021layered}), 
but its application to ground-based observations remains challenging due to the temporal evolution of the aberrations responsible for the speckle nuisance \citep{gerard2016planet, xie2022reference}.

To address these challenges, recent works have proposed learning a highly non-linear representation of the nuisance distribution without explicitly relying on similarity metrics or image subtraction.
\cite{wolf2024direct} leverage a large archival database from the Keck/NIRC2 instrument to build a generative model of the nuisance. 
An auto-encoder is trained to solve an inpainting task in a self-supervised fashion. 
The network's weights are optimized to maximize the fidelity of the reconstruction under a masked fraction of the field of view.
This approach uses patches with a large spatial context, allowing the network to capture the typical spatial structures of speckles.
A ridge regression step is performed on the residual images to exploit temporal correlations between consecutive images. 
Finally, flux and signal-to-noise (S/N) maps are produced using classical techniques based on image derotation, combination, and stacking. 
This approach performs better or on par with annular PCA.
\cite{chintarungruangchai2023possible} adapt a discriminative model of the nuisance using a residual learning technique. 
This model is built from the difference between low and high-quality images of the estimated flux distribution. 
Both image types are obtained through classical image derotation and stacking after nuisance attenuation by PCA decomposition. 
The low-quality image is derived from a subset of temporal frames used for the high-quality image, resulting in a higher S/N ratio in the high-quality image. 
The low-quality images serve as input to a CNN, which infers a high-quality image. 
The CNN's parameters are optimized by comparing the predicted high-quality image to the original ground truth.

\medskip

\textit{Our contributions:}
In this paper, we argue that information across observations of different targets can be leveraged to improve detection performance. 
(i) We propose a new supervised learning framework to improve detection performances of point-like sources:
by framing the detection problem as a reconstruction task, we demonstrate how this approach is well-suited for training an observation-independent model on a large database of observations. 
We build and apply the proposed approach on a database from the IRDIS (Infra-Red Dual-beam Imager and Spectrograph; \cite{dohlen2008infra}) of the SPHERE (Spectro Polarimetric High-contrast for Exoplanet Research; \cite{beuzit2019sphere}) instrument. 
(ii) Inspired by PACO (\cite{flasseur2018exoplanet}), which models correlations between pixels, we propose a new architecture that models the correlation between learned features. 
Similar to deep PACO (\cite{flasseur2024deep}), the architecture combines two complementary representations of the data: with either spatially co-aligned quasi-static speckles, or spatially co-aligned off-axis sources along the temporal dimension.
(iii) Furthermore, we describe a straightforward calibration procedure, enabling astronomers to deploy our calibrated model in practical scenarios.

We report an improved detection sensitivity at short angular separations, and find that our approach improves both the precision-recall trade-off and overall robustness over previous methods.

This paper is organized as follows. 
Section~\ref{sec:method} introduces the supervised learning framework of the proposed \thisalgo algorithm (for Multi-Observations DEep Learning model aided by COvariances). 
It also presents the detailed selection, preparation, curation and annotation of the database of multiple VLT/SPHERE observations leveraged by the proposed algorithm. 
In Sect.~\ref{sec:results}, we assess the performance of \thisalgo, in particular in terms of trade-off between precision and recall. 
Finally, Section~\ref{sec:conclusion} draws the conclusions of this paper.
We also included a glossary of technical terms in Appendix~\ref{app:glossary}.

%% file: method.tex
\section{Proposed method} 
\label{sec:method}

\begin{table}
	\centering
	\caption{Summary of the main notations.}
	\begin{tabular}{ccc}
		\toprule
		\textbf{Not.}\; & \textbf{Range}\; & \textbf{Definition}\\
		\midrule
		\multicolumn{3}{c}{$\blacktriangleright$ Constants}\\ 
		\midrule
		$T$ & $\mathbb{N}^*$ & number of temporal frames\\
		$H$ & $\mathbb{N}^*$ & height (pixels) of a frame\\
		$W$ & $\mathbb{N}^*$ & width (pixels) of a frame\\
		$h$ & $\mathbb{N}^*$ & height (pixels) of the PSF\\
		$w$ & $\mathbb{N}^*$ & width (pixels) of the PSF\\
		$K$ & $\mathbb{N}^*$ & number of pixels in a patch\\
		$L, C$ & $\mathbb{N}^*$ & number of features\\
		\midrule
		\multicolumn{3}{c}{$\blacktriangleright$ Data quantities}\\
		\midrule
		$\V y$ & $\mathbb{R}^{THW}$ & ADI sequence (measurements)\\
		$\V s$ & $\mathbb{R}^{THW}$ & nuisance component (speckles)\\
		$\V \alpha$  & $\mathbb{R}^{HW}$ & flux distribution (objects)\\
		$\V h$ & $\mathbb{R}^{hw}$ & off-axis PSF\\
		$\V m$ & $\mathbb{R}^{T(C)HW}$ & mask of known real sources\\
		$\V x$ & $\mathbb{R}^{HW}$ & spatial target signal linked to $\V \alpha$\\
		$\phi$ & $\mathbb{R}^T$ & vector of parallactic angles\\
		\midrule
		\multicolumn{3}{c}{$\blacktriangleright$ Operators}\\
		\midrule
		$\M R_\phi$ & $\mathbb{R}^{(C)HW} \mapsto \mathbb{R}^{T(C)HW}$ & multi-frame rotation\\
		$\M H$ & $\mathbb{R}^{(T)HW} \mapsto \mathbb{R}^{(T)HW}$ & spatial convolution with kernel $\V h$\\
		$\M N$ & $\mathbb{R}^{TCHW} \mapsto \mathbb{R}^{TCHW}$ & pixel-wise temporal normalization\\
    	$\text{MTA}_\phi$ & $\mathbb{R}^{CTHW}\mapsto\mathbb{R}^{THW}$ & masking \& temporal aggregation\\
    	\midrule
		\multicolumn{3}{c}{$\blacktriangleright$ Learning quantities}\\
		\midrule
      $\mathcal{D}_{\text{train}, \text{test}, \text{val}}$ & $ \{\V s_i \in \mathbb{R}^{THW}\}_i  $ & training, test, validation sets\\
		$f$ & $\mathbb{R}^{THW}\mapsto\mathbb{R}^{CTHW}$ & speckles-aligned learnable module\\
		$g$ & $\mathbb{R}^{CTHW} \mapsto \mathbb{R}^{HW}$ & objects-aligned learnable module\\		
		$\M F_{\V \theta} = g \circ f$ & $\mathbb{R}^{THW}\mapsto \mathbb{R}^{HW}$ & neural network parametrized by $\V \theta$\\
    	\midrule
		\multicolumn{3}{c}{$\blacktriangleright$ Estimated quantities}\\
		\midrule
		$\widehat{\V x}$ & $\mathbb{R}^{HW}$ & inferred spatial target signal\\
		$\widehat{\M S}$ & $\mathbb{R}^{K\times K}$ & empirical spatial covariance\\
		$\widehat{\M C}$ & $\mathbb{R}^{K\times K}$ & shrunk spatial covariance\\
		$\widehat{\rho}$ & $[0; 1]$ & shrinkage factor\\		
		\bottomrule
	\end{tabular}
	\label{tab:notations}
\end{table}

\noindent Let us now present the \thisalgo algorithm. 
We successively discuss the forward model of the observations (Sect. \ref{subsec:direct_model_observations}), 
the proposed supervised learning framework (Sect. \ref{subsec:supervised_learning}), 
our database of observations (Sect. \ref{subsec:database}), the construction of the training samples (Sect. \ref{subsec:dataset}), 
the architecture of the model (Sect. \ref{subsec:architecture}) and 
the implementation details (Sect. \ref{subsec:implementation_details}).

\begin{figure*}
		\includegraphics[width=\textwidth]{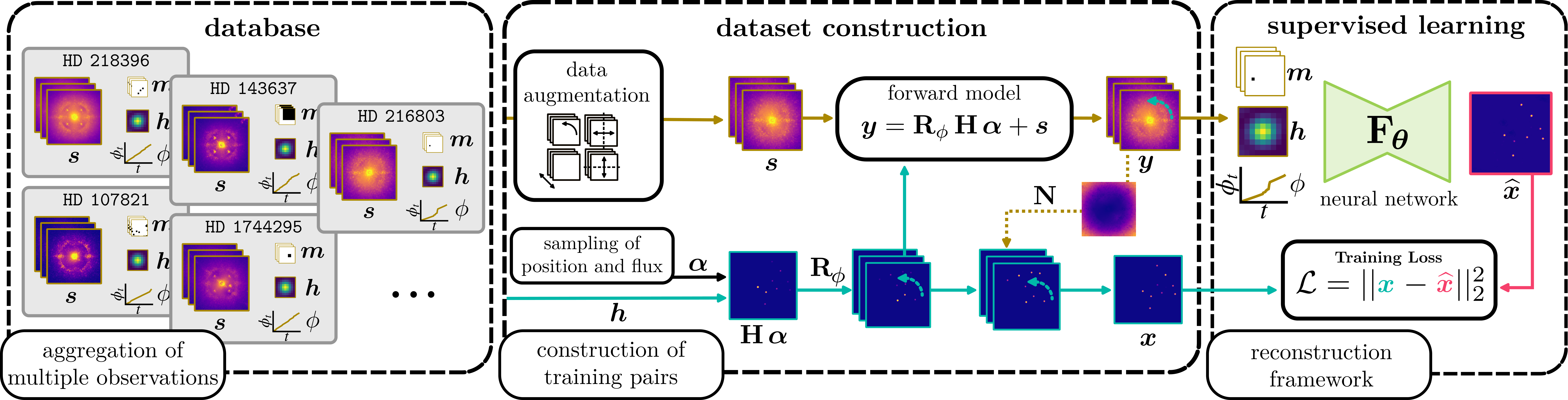}
    \caption{Overview of the proposed \thisalgo algorithm. \textbf{Left:} illustration of the database $\mathcal{D}$ containing multiple speckles realizations $\V s$ (shown in logarithmic scale), associated binary masks $\V m$ identifying known real sources, and vectors $\phi$ of parallactic angles. 
\textbf{Center:} construction of the training samples from observations drawn from $\mathcal{D}$. 
For a selected data cube, data augmentation is first applied, then synthetic sources are injected in $\V s$ through direct model (\ref{eq:data_model}). 
The synthetic signals are weighted by the variability of the speckles along their trajectories to form a target signal $\V x$ to be reconstructed. 
\textbf{Right:} reconstruction of the target signal $\V x$ by supervised learning from inputs $\V y$ and $\phi$. 
The weights $\V \theta$ of the learnable module $\M F$ are optimized by minimizing MSE between $\widehat{\V x}$ and $\V x$.
See text for details.
    }
    \label{fig:workflow}
\end{figure*}

Throughout the text, the reader can refer to Table \ref{tab:notations} summarizing the main notations, and to Fig. \ref{fig:workflow} illustrating the main steps of the proposed algorithm. 

\subsection{Direct model of the observations}
\label{subsec:direct_model_observations}

\noindent We consider the general case of $T$ individual exposures of $H\times W$ pixels recorded with the ADI technique using the pupil-tracking mode of the telescope. The resulting temporal images are calibrated and assembled (see Sect. \ref{subsec:database}) into a 3-D spatio-temporal data cube\footnote{For the sake of clarity, we denote multivariate data cubes as one-dimensional vectors.} $\V y \in \mathbb{R}^{THW}$ of intensity measurements. The latter is the result of two additive contributions through the following direct model:
\begin{equation}
	\V y = \M H \, \M R_\phi \, \V \alpha + \V s\,,
	\label{eq:data_model}
\end{equation}
where $\V \alpha \in \mathbb{R}^{HW}$ is the spatial flux distribution of the off-axis objects of interest, and $\V s \in \mathbb{R}^{THW}$ is the nuisance component (i.e., speckles and additive stochastic noise) corrupting signals of those sought objects. 
The forward operator $\M R_\phi: \mathbb{R}^{HW} \to \mathbb{R}^{THW}$ describes the (apparent) temporal rotation associated with the parallactic angles $\phi$ induced by ADI on the off-axis objects. 
In practice, this operator duplicates the input frame $T$ times, and each copy indexed by $t$ is then rotated by an angle $\phi_t$.
The operator $\M H: \mathbb{R}^{THW} \to \mathbb{R}^{THW}$ denotes the convolution operator with kernel $\V h \in \mathbb{R}^{hw}$, corresponding to the measured off-axis (non-coronographic) PSF. 
In this work, we consider solely unresolved point-like sources  (e.g., exoplanets, brown dwarfs, background stars) behaving as Dirac components within $\V \alpha$, 
and we left the reconstruction of spatially extended objects (i.e., circumstellar disks) for future work.

\subsection{Supervised learning strategy}
\label{subsec:supervised_learning}

\noindent We adopt a supervised learning framework, where we train a neural network $\mathbf{F}_{\V \theta}$ parametrized by weights $\V \theta$ to recover a \textit{target signal} $\V x$ built from the spatial distribution $\V \alpha$ of the sought sources (see Sect. \ref{subsec:dataset} for the link between $\V \alpha$ and $\V x$).
As real sources are scarce in direct imaging, we employ a semi-synthetic approach to build our training database: synthetic sources are generated using the measured off-axis PSF $\V h$, and injected into real measurements through the direct model (\ref{eq:data_model}). 
We denote by $\mathcal{D}$ the resulting database, from which we draw input measurements $\V y$, rotation vectors $\phi$, and target signals $\V x$.
As standard in the machine learning community, we split $\mathcal{D}$ into three non-overlapping subsets. 
We optimize the weights $\V \theta$ of a neural network over the training set $\mathcal{D}_{\text{train}}$, tune hyper-parameters in a validation step on the set $ \mathcal{D}_{\text{validation}}$, and finally evaluate the performance of the resulting model on the test set $\mathcal{D}_{\text{test}}$. 
These three datasets respectively represent 70\%, 15\%, and 15\% of the total number of samples in $\mathcal{D}$.

In our framework, we propose to cast the detection problem as a reconstruction task, where we minimize the error between a target signal $\V x$ and its estimate $\widehat{\V x}$ produced by the network.
The loss function used is the mean squared error (MSE). 
The network weights are thus estimated by \gls{empirical_risk_minimization} over $\mathcal{D}_{\text{train}}$:
\begin{equation}
  \min_{\V \theta} \ \sum_{(\V y,\, \phi,\, \V x) \sim \mathcal{D}_{\text{train}}} || \M F_{\V \theta}( \V y, \phi ) - \V x ||_2^2 ,
	\label{eq:no_condition}
\end{equation}
As detailed in Sect. \ref{subsubsec:pre_conditioning}, the target signal $\V x$ to recover is the signal of the source weighted by the variability of the speckles along its trajectory.
As such, the loss is proportional to the ``detectability'' of any given source.
Conversely, in deep PACO, all sources incur the same penalty, regardless of their relative difficulty (a binary similarity metric, adapted for very unbalanced classes, implemented as a Dice score).
In practice, we observed that training a neural network with loss (\ref{eq:no_condition}) is more stable than with a detection score, in particular when the quality of the observations is not optimal.
The proposed algorithm thus presents improved robustness to the heterogeneity of the database, as demonstrated empirically by our experiments in Sect. \ref{sec:results}.
In practice, the optimization of (\ref{eq:no_condition}) is performed on mini-batches of samples with \gls{stochastic_gradient_descent}.

\subsection{Database}
\label{subsec:database}

\begin{figure}
	\centering
  \begin{subfigure}{0.23\textwidth}
      \includegraphics[width=\textwidth]{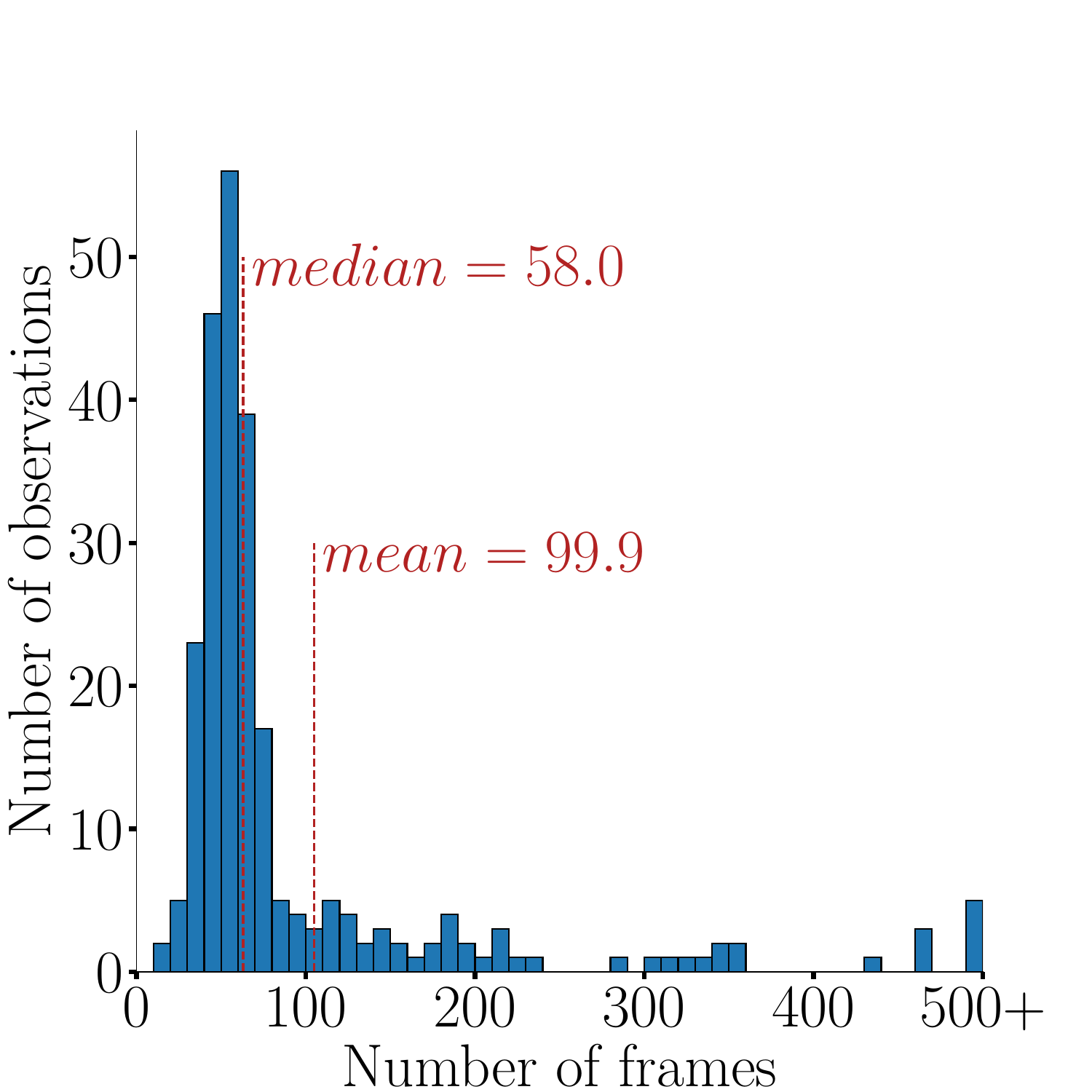}
      \caption{Number of frames}
      \label{fig:database_a}
  \end{subfigure}
  \begin{subfigure}{0.23\textwidth}
      \includegraphics[width=\textwidth]{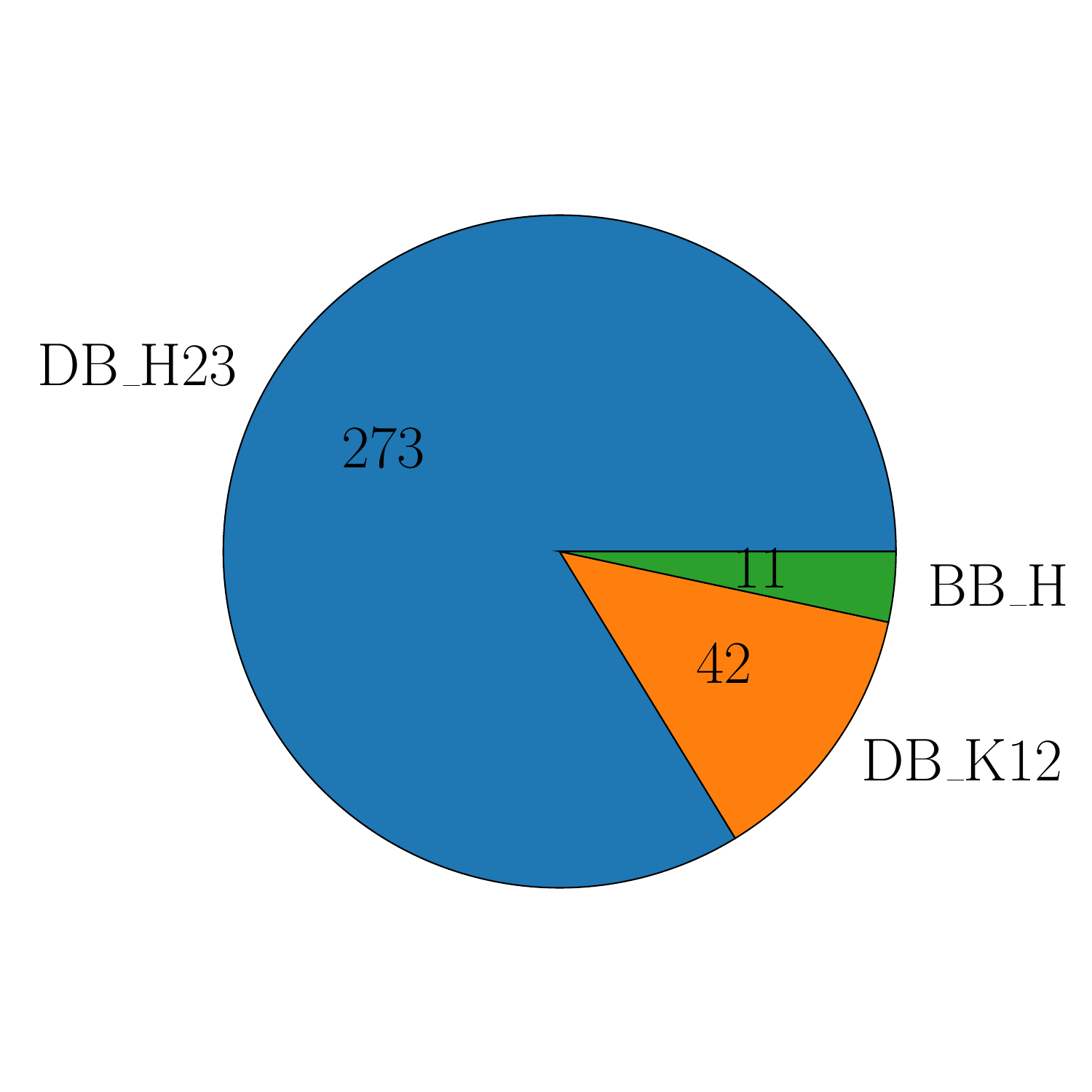}
      \caption{Infrared filters}
  \end{subfigure}
  \\
  \begin{subfigure}{0.23\textwidth}
      \includegraphics[width=\textwidth]{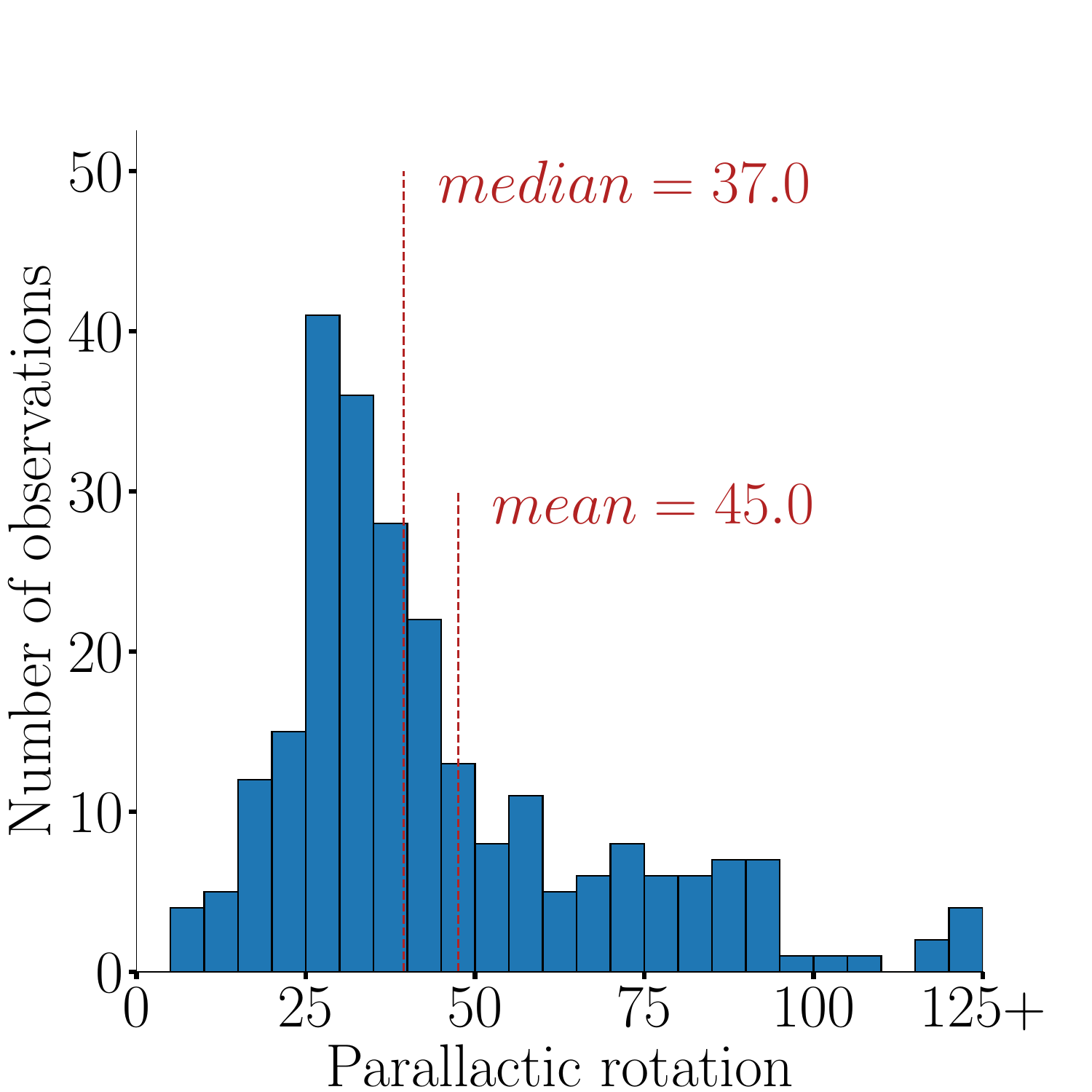}
      \caption{Parallactic rotations}
  \end{subfigure}
  \begin{subfigure}{0.23\textwidth}
      \includegraphics[width=\textwidth]{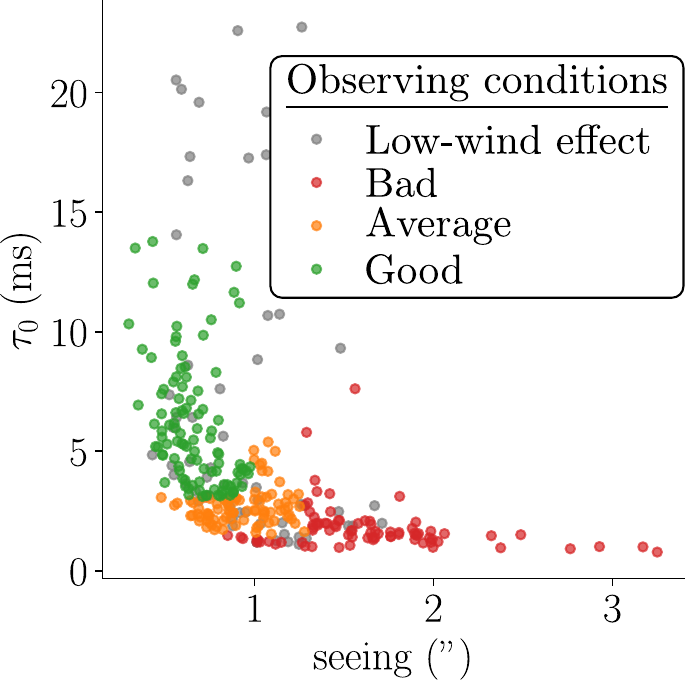}
      \caption{Observing conditions}
      \label{fig:database_d}
  \end{subfigure}
	\caption{Main statistics for the F150 database of observations of the VLT/SPHERE SHINE survey. (b): DB\_H23 is for observations conducted in the H2-H3 dual band ($\lambda_0 = 1.59 \, \micro \meter$, $\lambda_1 = 1.67 \, \micro \meter$), DB\_K12 stands for K1-K2 dual band ($\lambda_0 = 2.11 \, \micro \meter$, $\lambda_1 = 2.25 \, \micro \meter$), and BB\_H is for observations in broadband H ($\lambda \in \left[ 1.48, 1.77 \right] \, \micro \meter$). (d): The classification procedure between \textit{bad}, \textit{average}, and \textit{good} observing conditions based on seeing and coherence time $\tau_0$ is detailed in Sect. \ref{subsec:obs_cond}.
	}
	\label{fig:database}
\end{figure}

The nuisance component $\V s$ (dominated by speckles) is estimated via an archive of multiple ADI observations from the SPHERE-IRDIS instrument. 
The database used in this paper is the so-called F150 subset of the SHINE large survey designed for exoplanet imaging \citep{desidera2021sphere, langlois2021sphere, vigan2021sphere}. 
The full F150 archive is comprised of 322 individual data cubes recorded with diverse parallactic rotations, numbers of frames, and spectral filters (Fig. \ref{fig:database}). 
The associated observing conditions are quite heterogeneous as the survey has been conducted over several years.  

The raw observations are pre-reduced with the data reduction and handling pipeline (DRH; \cite{pavlov2008sphere}) of the SPHERE instrument, which performs thermal background subtraction, flat-field correction, anamorphism correction, compensation for spectral transmission, flux normalization, bad pixels identification and interpolation, true-North alignment, wavelength calibration, astrometric calibration, and frame selection. 
These operations are then complemented by a custom calibration implemented in the SPHERE data center \citep{delorme2017sphere}, in particular to improve bad pixels correction. 
Finally, the resulting temporal frames are precisely centered making use of replicas of the star (which is hidden by the coronagraph), namely \textit{waffle satellite spots}, created by introducing a 2-D sinusoidal pattern on the high-order deformable mirror \citep{beuzit2019sphere}. 
Fitting 2-D anisotropic Gaussian models on the four satellite spots and interpolating images allows to reach an overall centering accuracy of about 0.1 to 0.05 pixel, thus maximizing signals recombination of the sought objects with ADI  \citep{chomez2023preparation,dallant2023pacome}. 
When satellite spots are absent (or present only on a few frames at the beginning and at the end of the observation sequence) we rely only on the pointing stability of the instrument.

We curate the database manually to discard irrelevant observations.
We exclude all observations of stars hosting spatially extended structures such as circumstellar disks, or with a cluttered field of view (too many background stars for example).
We also discard observations having less than $T=16$ frames because the temporal diversity is insufficient to extract adequate statistics about the speckles.
On the remaining observations, we flag known sources based on detection results reported in \cite{langlois2021sphere}, that were obtained with the TLOCI and KLIP-PCA detection algorithms.
To the best of our knowledge, this is the most comprehensive reference regarding known sources (exoplanets, background stars, brown dwarfs) in the F150 archive to date.
We also \textit{manually} identify and flag additional (candidate) point-like sources based on additional reductions of the data performed with the PACO algorithm.
Using PACO, we typically flag all point-like sources detected at a S/N higher than 5 in a single epoch and sources consistently detected at multiple epochs with a S/N higher than 4.
Given these detection thresholds and the considered number of observations, we expect to experience a few dozen false alarms on the whole database. 
This does not limit our approach because discarding portions of data associated with false alarms only slightly reduces the available training data volume. 
Conversely, keeping some real sources (not detected with KLIP, TLOCI, PACO, or mislabeled) is not critical as we resort to a data-augmentation strategy at training time (see Sect. \ref{subsec:dataset}).
The augmentation scheme involves a temporal shuffling of the frames, which suppresses the temporal signal consistency of any real sources.
Although not essential, annotating known sources helps prevent instability during training caused by bright sources (e.g., background stars). 
Indeed, such sources can be masked out during the forward pass of the network, as detailed in Sect. \ref{subsubsec:pre_conditioning}.

In its dual-band mode, the IRDIS imager of SPHERE, captures two images simultaneously from closely spaced spectral channels: e.g.  H2 ($\lambda = 1.59 \, \micro\meter$) and H3 ($\lambda = 1.63 \, \micro\meter$). 
As we need a sufficient number of training samples with similar speckles, we consider only observations from the H2-H3 dual band, which is the most represented in the F150 subset, see Fig. \ref{fig:database}(b).
We keep both spectral channels in our database.
If needed and without loss of generality, several models can however be trained separately on each spectral channel, whenever the number of available training samples is sufficient, see Sect. \ref{subsec:ablation} for an ablation study. 
As we consider ADI, we are by definition in mono-spectral mode. 
We left for future work the problem of jointly processing spatio-temporo-spectral data cubes recorded with ASDI. 

After this data curation and selection process, we end up with 220 annotated data cubes from the H2-H3 dual band, which represent 20,462 individual frames in total. 
Each data cube is paired with a corresponding mask cube $\V m \in \mathbb{R}^{THW}$ delimiting pixels affected by known real sources.

\subsection{Construction of the training samples}
\label{subsec:dataset}

\noindent As explained in the rest of this section, we create pairs of semi-synthetic measurements for creating pairs of measurements $\V y$ and ground truth signals $\V x$.
We discuss here, the astro-photometric parameters of the simulated sources added to the training measurements, the data augmentation strategy we employ, and the pre-conditioning of the data applied in order to train a CNN from samples corrupted by a strong and spatially non-stationary nuisance.

\subsubsection{Extraction of clips}
\label{subsubsec:clips}

We focus on sources at short angular separation, i.e. the area where the discrepancy between the ultimate detection capability driven by the fundamental photon noise limit and the achievable contrast is the highest, see \cite{flasseur2024deep}. 
In practice, we crop data cubes such that $H=W=256$ pixels, corresponding to an inscribed circle with a radius of 1.56 arcseconds.

As illustrated by Fig. \ref{fig:database}(a), the number of frames can vary significantly from one observation to the other. 
However, to allow the training of our model with mini-batches, it is necessary for the inputs to have a fixed size.
To achieve this,  we extract segments of consecutive frames for each observation in the database, referred to as \textit{clips}, with fixed temporal length $T_\text{clip}$. 
Each clip is paired with a corresponding mask and rotation vector, extracted in a similar manner.
For observations with $T$ frames where $T < T_\text{clip}$ frames, we extract only one clip padded with zeros.
The corresponding binary mask is filled with zeros in the padded area, ensuring that this part will be ignored in the forward pass of the neural network,
as detailed hereafter.
We apply this procedure for each spectral channel (H2 and H3), resulting in a collection of mono-spectral clips.

\subsubsection{Position and flux sampling}
\label{subsubsec:position_flux_sampling}

At training time, clips are injected with sources in a circular area with a radius of 1.38''.
For each sample in $\mathcal{D}$, we draw a number of synthetic sources within $[\![3, 10  ]\!]$.
This is a realistic scenario as we expect a few faint point-like sources in the field of view.
This also allows for the consideration of several sources per training sample, which is critical for speeding up the optimization process.
The initial positions of the sources on the first temporal frame serving as a reference, are drawn uniformly.
The influence of the sampling distribution was studied in our previous work \cite{flasseur2024deep}, and we have found very similar detection performance between uniform sampling in cartesian and polar coordinates.
Given the typical spread of the off-axis PSF, we enforce that synthetic sources are at least 25 pixels away from each other to avoid overlap.
We also make sure that synthetic sources do not overlap with known real sources identified with the procedure described in Sect.
\ref{subsec:database}.

Once the positions of the synthetic sources are determined, we sample the relative flux $\alpha_i$ of each source $i$ as:
\begin{equation}
  \alpha_i = \frac{\kappa \, \sigma_i \, u_i}{\sqrt{T} \, ||\V h||_\infty} \,, \quad u_i \sim \mathcal{U}(0, 1)\,,  
  \label{eq:flux_synthetic_sources}
\end{equation}
where $\sigma_i$ is the standard deviation (along the temporal dimension) of the nuisance $\V s$ under the source at its initial position,
and $||\V h||_\infty$ the amplitude of the off-axis PSF.
Equation \ref{eq:flux_synthetic_sources} allows to roughly control the level of difficulty of injected sources with the hyper-parameter $\kappa$, set to $\kappa = 4$ in practice.
This typically prevents the sampling of bright sources in ``easy'' regions (those with low variability of speckles), which are not of interest as they are easily detectable.

\subsubsection{Data augmentation}
\label{subsubsec:data_augmentation}

Before adding the synthetic objects and the nuisance component $\V s$ through the direct model (\ref{eq:data_model}), we perform data augmentation on $\V s$ and $\phi$ during training including temporal shuffling of the frames; horizontal, vertical and temporal random flipping as well as random spatial rotations by 90 degrees.
Besides, the rotation vector $\phi$ is also randomly flipped and reversed (multiplication by $-1$).
Data augmentation is a way to artificially expand our database by leveraging prior knowledge of the problem.
It also ensures that unknown (or misflagged) real sources become inconsistent with the forward model, and do not hinder training by being considered as false positives.
During validation and evaluation, we only flip the rotation vector.
This ensures that the conclusions are not flawed by mislabeled real sources, while keeping the frames in their original order and orientation.

\subsubsection{Pre-conditioning}
\label{subsubsec:pre_conditioning}

Naively training a neural network to directly recover the raw signal of the sources $\M H \V \alpha$ from measurements of variable temporal lengths is doomed to fail.
Indeed, the loss incurred must be commensurate with the detectability of a source, otherwise, the neural network training becomes unstable.
The two primary factors affecting the detectability of sources are:

\textbf{High contrast and variability of the speckles:}
        The speckles in the inner region are several hundred times brighter than those in the outer region.
        Additionally, their amplitude and variability are higher in this region.
        Consequently, a source with a given flux will be much more difficult to recover at short angular separations than farther away.

\textbf{Number of effective frames:}
        The number $T$ of temporal frames can drastically change across observations.
        For similar targets, observing conditions and instrumental settings, the off-axis sources are easier to detect with more frames in an observation.
        Besides, for each pixel in the object space, the number of observed data points varies. 
        Indeed, because the objects rotate and the field of view is not circular, some pixels in the object space are only visible in a subset of frames in $\V y$.
        Their effective number of frames is thus inferior to $T$.
        Moreover, areas with known sources need to be discarded during training, which also leads to a spatial variation of the effective number of frames within an observation.

\medskip

\noindent Consequently, it is necessary to pre-condition the signal of the injected sources to account for these two effects.
We propose to pre-condition the problem by applying specific transformations to the object signal $\M H \, \V \alpha$, in order to obtain the target signal $\V x$.
Symmetrically, these transforms are also applied within the neural network, to obtain the estimate $ \widehat{\V x}$ of the target signal $\V x$. This pre-conditioning is formed by the two following steps:

\textbf{Speckles normalization:}
        In order to neutralize the effect of the spatially-varying intensity of the speckles, we need to take into account the pixel-wise standard deviation of the speckles.
        We denote $\M N$ the normalization operator defined by: 
        \begin{align}
        	\M N: \mathbb{R}^{THW}  &\to \mathbb{R}^{THW}\,,\nonumber\\
           (\V z_0, \cdots, \V z_{t-1} ) &\to (\V z_0 \circ \V n, \cdots, \V z_{t-1} \circ \V n )\,,
        \end{align}
        \noindent with $\circ$ the \gls{hadamard_product}, and $\V n \in \mathbb{R}^{HW}$ the pixel-wise inverse of the measurement standard deviation (along the temporal dimension). Formally, $\V n$ is defined with:
        \begin{align}
          \V n_j = \sqrt{\frac{T-1}{\sum_{t=0}^{T-1} (\V y_{t,j} - \overline{\V y}_{j})^2}}\,, \quad \forall j \in [\![0, HW -1  ]\!]\,.
        \end{align}
and $\overline{\V y} \in \mathbb{R}^{HW}$ the pixel-wise temporal mean of the measurement:
  \begin{align}
  \overline{\V y}_j = \frac{1}{T} \sum_{t=0}^{T-1} \V y_{t,j}\,, \quad \forall j \in [\![0, HW -1  ]\!]\,.
  \end{align}

\begin{figure}
	\centering
	\includegraphics[width=.485\textwidth]{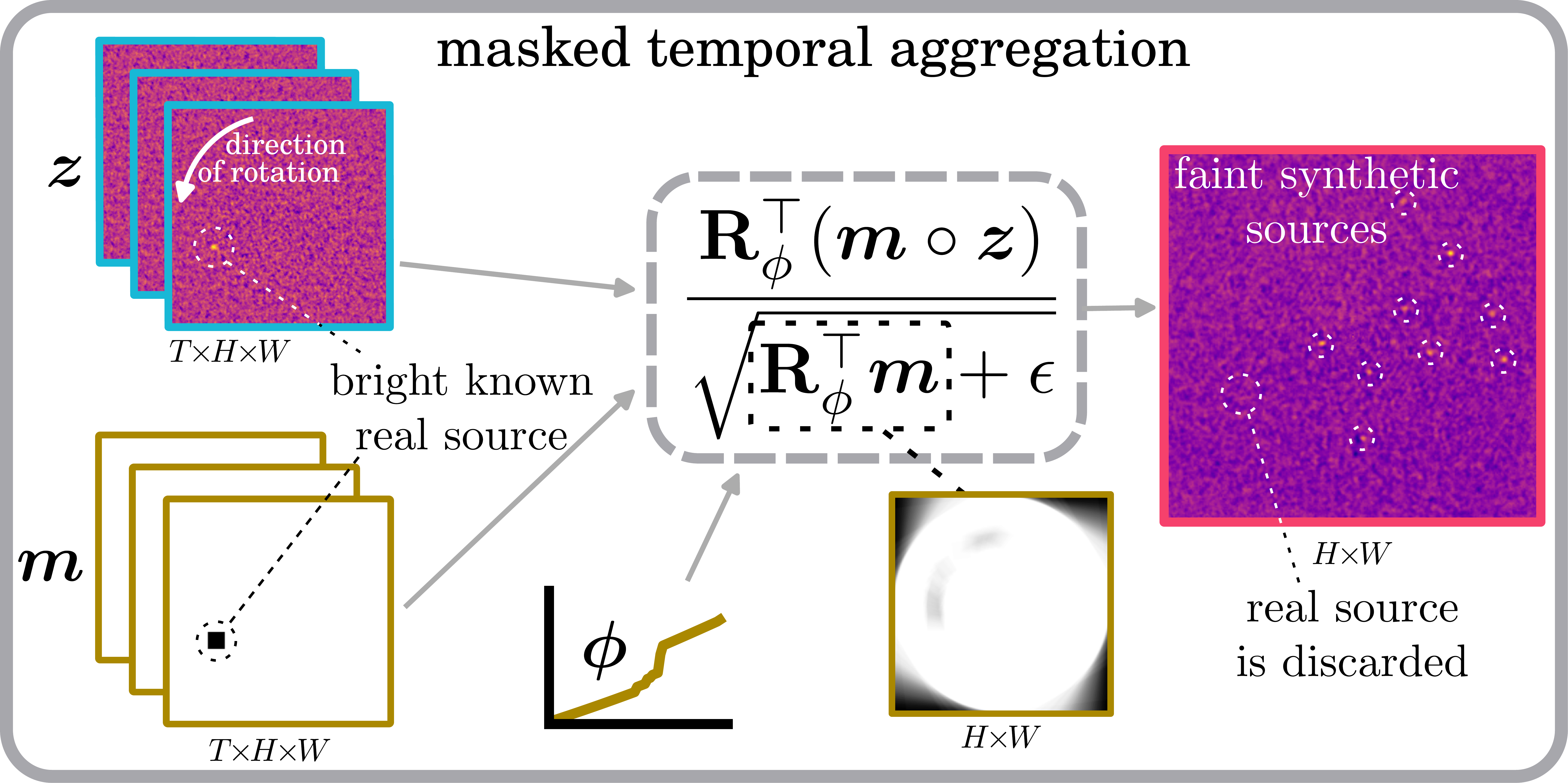}
	\caption{Schematic representation of the masked temporal aggregation operator (MTA) applied for pre-conditioning of the training samples. This procedure corresponds to the grey box in the architecture of the proposed method given in Fig. \ref{fig:architecture}.}
	\label{fig:mta}
\end{figure}

\textbf{Masked temporal aggregation:}
	We propose a simple \textit{masked temporal aggregation} (MTA) operator, to tackle the spatial variation of the number of effective frames.
        The MTA operator is defined as:
        \begin{align}
          \text{MTA}_\phi : \mathbb{R}^{CTHW} \times \{0, 1\}^{THW} &\to \mathbb{R}^{CHW}\nonumber\\
          (\mathbf{z}, \V m) &\to \frac{\mathbf{R}_\phi\T (\V m \circ \V z)}{\sqrt{\mathbf{R}_\phi\T \V m + \epsilon}}\,,
          \label{eq:mta}
        \end{align}
        \noindent where $\V m \in \{0, 1\}^{THW}$ denotes a mask accounting for valid pixels, i.e., 
        pixels unaffected by known sources, located within the field of view, and part of valid temporal frames.
        The operator $\M R_\phi\T$ rotates each frame $t$ by an angle $-\phi_t$ to align the sources, and then sums the aligned frames
        \footnote{We slightly overuse the notation here: if $C > 1$, the mask $\V m$ is duplicated to match the dimension of $ \V z$ in the numerator}.
        The quantity $\M R_\phi\T \V m \in \mathbb{R}^{HW}$ in the denominator represents the number of valid pixels encountered by an object initially located at that point.
        The MTA operator allows to handle inputs $\V z$ of variable temporal length.
        If $\V z$ is a vector of \gls{iid} random variables, then the variance of the output $MTA(\V z, \V m)$ is fixed and spatially uniform, regardless of the number of frames $T$, or mask $\V m$.
        As discussed in Sect. \ref{subsec:architecture}, this property is useful since we can then apply a denoiser with a fixed noise level for all observations.
        The MTA operator also handles border effects, as the mask is set to zero outside the field of view.
        The scalar $\epsilon$ is introduced in Eq. (\ref{eq:mta}) for numerical stability, because the number of effective frames may be zero in some situations (e.g., the presence of a known source at very short separation thus leading to a spatially quasi-stationary masking along the temporal dimension). 
        This parameter is set to $10^{-3}$ in practice.
        The MTA operator is represented in Fig. \ref{fig:mta}.

\medskip

\noindent Finally, we combine speckles normalization with MTA to obtain the target signal to be reconstructed:
\begin{equation}
  \V x =  \text{MTA}_\phi (\M N \, \M R_\phi \, \M H \, \V \alpha, \V m)\,.
  	\label{eq:target_signal}
\end{equation}
Analogous transformations are also performed within the neural network architecture, as described in Sect.~\ref{subsec:architecture}.

\subsection{Model architecture}
\label{subsec:architecture}

\begin{figure}
	\centering
	\includegraphics[width=.48\textwidth]{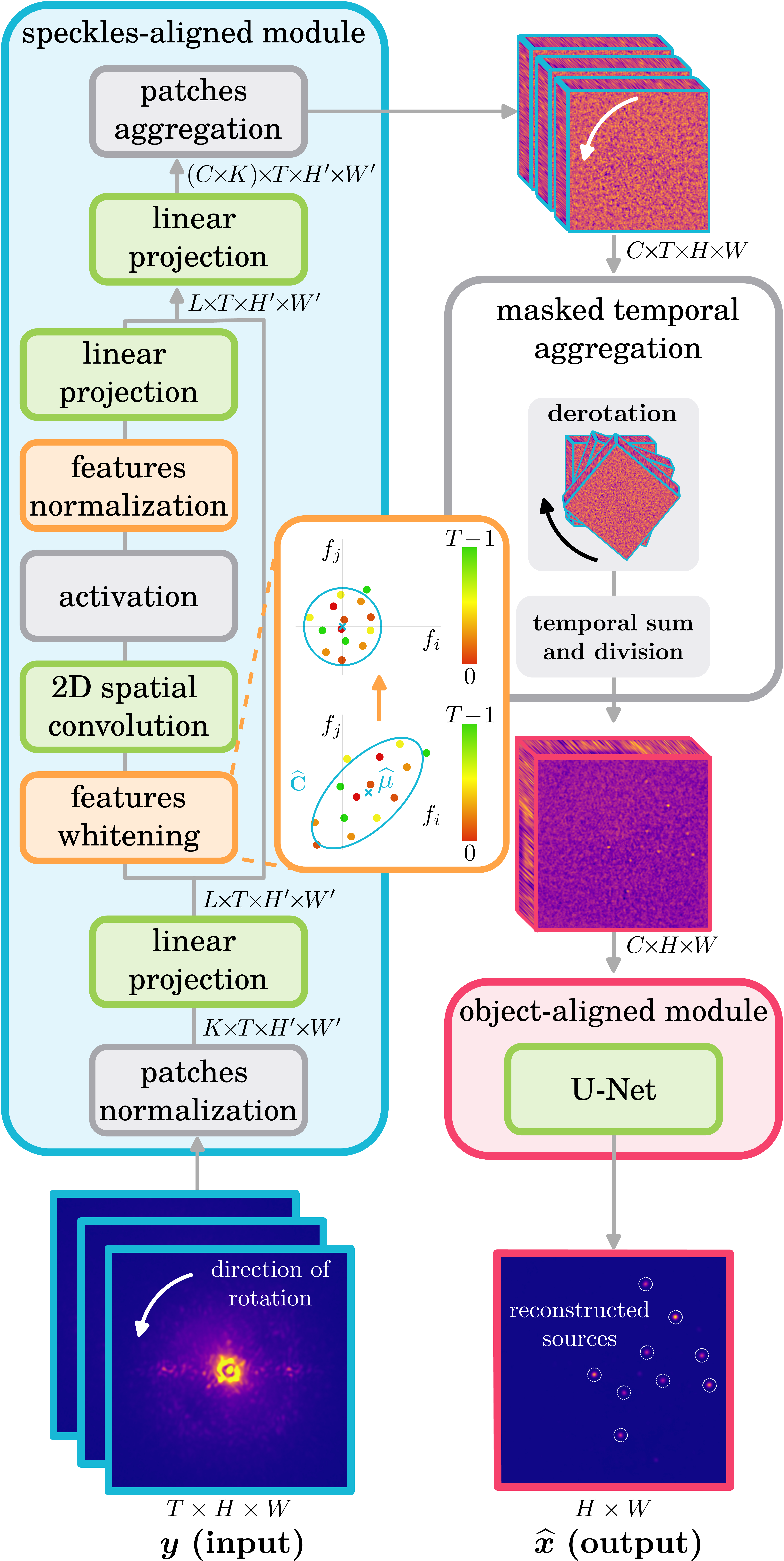}
	\caption{
Schematic representation of the architecture of \thisalgo.
The first processing stage of the proposed approach, working on speckles-aligned images, is depicted in the blue box.
The sub-blocks associated with the whitening and normalization of the features through patch covariances are in orange (the central part illustrates the effect of the whitening on two arbitrary features $(f_i, f_j)$), sub-blocks containing learnable parameters (projections and 2-D spatial convolutions) are highlighted in green, and other operations (non-linear activation, patch extraction, normalization and aggregation are in grey).
The output of the first stage is derotated and temporally aggregated by the MTA block.
For illustration purposes, the rotation vector and the binary mask are omitted in this figure.
A more detailed view of the MTA module is represented in Fig. \ref{fig:mta}.
The object-aligned features are then filtered by a \gls{unet} to produce the final reconstruction $\widehat{\V x}$.
For each stage, examples of some intermediate quantities and their associated shapes are given.
}
	\label{fig:architecture}
\end{figure}

Our model takes as in input a measurement cube $\V y \in \mathbb{R}^{THW}$, and outputs a 2-D reconstruction $\widehat{\V x} \in \mathbb{R}^{HW}$.
Its architecture is composed of two stages: speckles-aligned and object-aligned stages working on two complementary views of the data.
A key aspect of our architecture is that both the speckles-aligned and the object-aligned blocks are trainable, contrary to state-of-the-art exoplanet detection algorithms based on \gls{deep_learning} where only the object-aligned stage is, see Sect. \ref{sec:introduction}. We describe in Sects. \ref{subsubsec:speckles_aligned_stage} and \ref{subsubsec:object_aligned_stage} the rationale and content of these two blocks, which are demonstrated by Fig. \ref{fig:architecture}.

\subsubsection{Speckles-aligned stage}
\label{subsubsec:speckles_aligned_stage}

At the speckles-aligned stage, overlapping square patches of $K$ pixels are extracted with a stride of $\sqrt{K} / 2$.
After this step, the dimension of the data is $K\times T \times H' \times W'$, where $H'\times W'$ represents the total number of stacks of patches. 
Then, each temporal stack of patches is centered around its mean and normalized by its standard deviation.
This step is necessary to partly mitigate the high contrast of the speckles. 
It is analogous to the normalization with operator $\M N$ described in Sect.~\ref{subsec:dataset}, with an additional centering.
This coarse normalization ensures that patches are roughly centered and rescaled.
Each normalized patch of size $K$ is then linearly projected into a higher dimensional feature space of dimension $L$,
resulting in a feature map of shape $L\times T \times H' \times W'$.

We adopt a \gls{residual} architecture to process extracted features, as is standard for image-to-image architectures.
Similar to U-Net residual blocks, the residual branch is comprised of linear, normalization, and non-linear modules.
A key element of our model is the normalization module, which we describe thereafter.

\textbf{Normalization:}
In the computer vision community, various types of \glspl{normalization_layer} have been proposed over the last few years, most notably batch \citep{ioffe2015batch},
layer \citep{ba2016layer}, instance \citep{ulyanov2016instance} and group \citep{wu2018group} normalization.
Normalization is known to improve the convergence rate and generalization performance of deep neural networks \citep{bjorck2018understanding}.
Some works go beyond a simple centering and re-scaling of the features, and take covariances into account by whitening the features \citep{huang2018decorrelated, huang2019iterative}, but do not apply any regularization to the estimated parameters accounting for finite sampling.

In the direct imaging community, modeling spatial covariances between neighboring pixels has proven crucial to improve detection performance \citep{flasseur2018exoplanet,flasseur2020paco,flasseur2020robustness}.
For instance, the PACO algorithm infers, in the maximum likelihood sense, spatial covariances $\widehat{\M S} \in \mathbb{R}^{K\times K}$ within patches. However, the sample covariances $\widehat{\M S}$ are very noisy and can be rank-deficient due to the limited number $T$ of frames available to perform the estimation. 
They are thus regularized by \gls{shrinkage} that consists in a \gls{convex_combination} of the high-variance/low-bias estimator $\widehat{\M S}$ with a low-variance/high-bias regularization matrix $\widehat{\M Q}$ having fewer degrees of freedom:
\begin{equation}
  \widehat{\M C} = (1 - \widehat{\rho}) \, \widehat{\M S} + \rho \, \widehat{\M Q}\,,
  \label{eq:shrinkage_convex_combination}
\end{equation}
with $\widehat{\rho} \in [0; 1]$ a key hyper-parameter striking a \gls{bias_variance_tradeoff}. 
The deep PACO algorithm combining a statistical model of the nuisance with a learnable model also leverages the shrunk covariances $\widehat{\M C}$ during a pre-processing step via patch whitening.
The resulting frames are then derotated and fed to a U-Net \citep{ronneberger2015u} to perform source detection by solving a pixel-wise binary classification task.

In this paper, we propose to bridge the gap between both communities, and apply shrinkage to covariances of learned features.
To reduce memory usage, features are split into $G$ groups, such that $L = G \times L'$.
Groups of features are independently whitened and then concatenated afterwards.
This is especially important since the memory footprint of the covariance matrices scales in $\mathcal{O}(L^2)$.
For a given group, the whitening procedure is as follows.
Let $\mathbf{X} \in \mathbb{R}^{TL'}$ be a stack of $T$ features of dimension $L'$.
First, features are centered:
\begin{equation}
  \mathbf{X}_c = \left( \M I_T - \frac{1}{T} \M 1_T \right) \, \mathbf{X}\,.
\end{equation}
Then, the sample covariance matrix $\widehat{\M S}$ is computed:
\begin{equation}
  \widehat{\M S} = \frac{1}{T} \M X_c\T \, \M X_c\,. 
\end{equation}
As in PACO, we consider a non-uniform regularization matrix $\widehat{\M Q}$ accounting for the sample variances: $\widehat{\M Q}_{i,j} = 0$ if $i \neq j$ and $\widehat{\M Q}_{i,i} = \widehat{\M S}_{i,i}$ otherwise. With these specific forms for $\widehat{\M S}$ and $\widehat{\M Q}$, the shrinkage factor $\widehat{\rho}$ can be estimated in a data-driven fashion via the following closed-form expression:
\begin{equation}
  \widehat{\rho} = \frac{\tr( \widehat{\M S}^2 ) + \tr^2(\widehat{\M S}) - 2 \tr( \widehat{\M S} \circ \widehat{\M S} )}{( T + 1 ) ( \tr( \widehat{\M S}^2) - \tr( \widehat{\M S} \circ \widehat{\M S} ) )}\,,
  	\label{eq:shrinkage_rho}
\end{equation}
where $\text{tr}$ is the \gls{trace} operator. 
The estimated regularization amount $\widehat{\rho}$ approximates the optimal setting minimizing the estimation risk between the shrunk covariance and the true (but unknown) covariance \citep{flasseur2024shrinkage}. As the model of the covariance is local, the shrinkage quantity (\ref{eq:shrinkage_rho}) also adapts to the non-stationarities of the data. 
Given the shrunk covariance $\widehat{\M C}$, features are whitened with the \gls{cholesky_factorization} of the associated precision matrix:
\begin{equation}
	\mathbf{X}_w = \mathbf{X}_c \, \widehat{\M W} \text{~~such that~~} \widehat{\M W} \, \widehat{\M W}^\top = \widehat{\M C}^{-1}\,.
	\label{eq:whitening}
\end{equation}

\textbf{Other modules:} 
After data normalization by whitening, we perform 2-D spatial convolution, followed by a non-linearity.
We then apply a feature-wise temporal normalization (centering and re-scaling).
Finally, we perform a linear projection.
The resulting (transformed) features are added to the input features.

\textbf{Feature aggregation:}
After the residual block, features are projected back to their original measurement shape.
To do so, we perform a linear projection to transform stacks of features of dimension $L \times T \times H' \times W'$ to stacks of patches of dimension $ (C \times K) \times T \times H' \times W'$.
We found the best performances for $C=4$.
Patches are then aggregated to retrieve the same spatial and temporal dimensions as the measurement $\V y$.
As this stage, the data cube is of dimension $C \times T \times H \times W$.

In the following, we denote by $f_{\V \theta_1}: \mathbb{R}^{THW} \rightarrow \mathbb{R}^{CTHW}$ the learnable speckles-aligned module.

\subsubsection{Object-aligned stage}
\label{subsubsec:object_aligned_stage}

In order to temporally aggregate the output of the speckles aligned stage, we make use of the MTA operator described in Sect.~\ref{subsec:dataset}.
At this stage, the features have been centered and whitened, we can consider the background noise to be mostly uniform and uncorrelated.
As previously mentioned, under these conditions, the noise level after the MTA operator is fixed and spatially uniform.

We can thus apply a learned denoiser $g_\theta$, which produces the final output.
The speckles and object aligned blocks can be summarized as:
\begin{equation}
  \V x = \mathbf{F}_{\V \theta}( \V y, \phi) = g_{\V \theta_2}\left( \text{MTA}_\phi\left(f_{\V \theta_1} (\V y),  \V m \right) \right)\,,
\end{equation}
\noindent where $\V \theta = [\theta_1, \theta_2]$ is the combined weights of the network, and $\V \phi$ is the parallactic rotation vector.

\subsubsection{Model ensembling}
\label{subsucsec:model_ensembling}

We resort to \textit{model ensembling} to produce the final reconstruction $\widehat{\V x}$. 
In deep learning, this refers to the practice of combining individual predictions from multiple models to produce a final prediction \citep{dong2020survey},
and is based on the idea that different models may capture different aspects or patterns within the data. 
This usually increases the performance and robustness of the model by reducing over-fitting and improving generalization \citep{ganaie2022ensemble}. 
There are several ways to obtain the diversity needed to perform model ensembling \citep{dong2020survey}. 
In this work, we use a standard approach based on bootstrap aggregation that consists in training multiple instances of the same model, with different random seeds.
Both the weights initialization, the order of the training data and the random parameters defining training sources differ from one model to the other. 
In our case, we have found empirically that this approach is useful, in particular to mitigate false alarms that can be hallucinated by the model.
In practice, we average $M=10$ reconstructions from different models, and we have experienced no significant additional gain by considering more models, see Sect. \ref{subsec:ablation} for an ablation analysis. 

\begin{figure}
	\centering	
	\includegraphics[width=.48\textwidth]{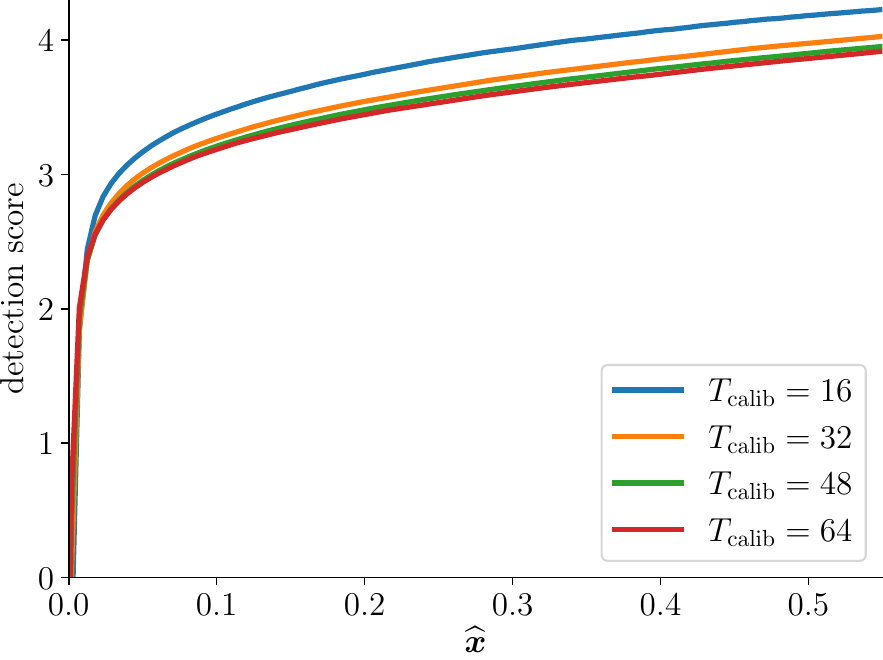}
    \caption{Calibration mapping $c$ linking the natural output values $\widehat{\V x}$ of \thisalgo to a detection score interpretable as a S/N of detection. The experiments are conducted with different numbers $T_{\text{calib}}$ of frames within the calibration datasets in the absence of sources of interest.
  }
    \label{fig:calib_n_frames}
\end{figure}

\subsection{Calibration}
\label{subsec:calibration}

In the previous section, we have shown how the detection problem can be framed as a reconstruction problem.
However, the network output $\widehat{\V x}$ is not directly interpretable in terms of probability of false alarm (PFA).
Therefore, a calibration step is necessary to tie both quantities.
To do so, we need to characterize the output distribution of our method under the null-hypothesis $\mathcal{H}_0$ that no source is present.
We can then use a simple hypothesis testing framework to obtain the probability of false alarm from a given pixel value.

\subsubsection{Additional assumptions}
\label{subsucsec:calibration_assumptions}

We denote the output distribution under the null hypothesis as
  $\widehat{\V x}_i(\V y) | \mathcal{H}_0$,
  with $\V y$ the measurement (with its rotation vector, omitted here for brevity),  $\widehat{\V x}(\V y) \in \mathbb{R}^{HW}$ the output of our method, and $i$ the pixel index.
We make a few additional assumptions regarding this distribution:
\begin{enumerate}
  \item \textbf{spatial stationarity --} the output distribution is identical for all pixels under the null-hypothesis:
    \begin{equation}
      \forall \, \V y \sim \mathcal{D}, 
      \forall \ (i, j) \in [0, HW-1]^2, \quad 
      \widehat{\V x}_i(\V y) | \mathcal{H}_0 \, \eqd \, \widehat{\V x}_j(\V y) | \mathcal{H}_0\,;
    \end{equation}
    \noindent where $\eqd$ denotes the \gls{equality_distribution}.
  \item \textbf{observation independence --} the output distribution is the same for all  $ \V y \sim \mathcal{D}$,
  with $\mathcal{D}$ the distribution of observations:
    \begin{equation}
      \forall \, (\V y, \V y') \sim \mathcal{D}, 
      \forall \ i \in [0, HW - 1], \quad  
      \widehat{\V x}_i(\V y) | \mathcal{H}_0 \, \eqd \, \widehat{\V x}_i(\V y') | \mathcal{H}_0\,;
    \end{equation}
  \item \textbf{pixel independence --} we assume pixels are independent, i.e., we ignore spatial correlations in the output maps:
    \begin{multline}
      \forall \, \V y\sim \mathcal{D}, 
      \forall \ (i, j) \in [0, HW-1]^2, \quad 
      \widehat{\V x}_i(\V y) | \widehat{\V x}_j(\V y), 
      \mathcal{H}_0 \, \eqd \, \widehat{\V x}_i(\V y) | \mathcal{H}_0\,.
    \end{multline}
\end{enumerate}
Assumption (iii) is made instead of reasoning in terms of spatially correlated detection blobs for simplicity (the latter often requires a dedicated procedure to identify individual blobs and remove iteratively the associated source contributions within the data, see for instance \cite{flasseur2018unsupervised,flasseur2021finding}). 
Both strategies are essentially equivalent as a single detection blob typically produces the same amount of multiple pixel detections in the case of a false alarm than in the case of a true detection.
We emphasize that these assumptions are applied to the output distribution of the network, following feature whitening, which is intended to remove correlations between pixels affected by speckles. 
While idealized, these assumptions have proven to work effectively in practice.

In this framework, we thus note the output distribution under the null hypothesis $\widehat{\V x} | \mathcal{H}_0$ such that:
\begin{equation}
  \forall \, \V y\sim \mathcal{D}, 
  \forall i \in [0, HW-1 ], 
  \quad \widehat{\V x}(\V y)_i | \mathcal{H}_0 \, \eqd \, \widehat{\V x} | \mathcal{H}_0\,.
\end{equation}

\subsubsection{Calibration set}
\label{sec:calibration_set}

In order to characterize the distribution of $\widehat{\V x} | \mathcal{H}_0$, we resort to a calibration set $\mathcal{D}_{\text{calib}}$.
This calibration set is comprised of 15 observations, in the H2-H3 spectral bands, that are not part of the sets $\mathcal{D}_{\text{train}}, \mathcal{D}_{\text{val}}$ or $\mathcal{D}_{\text{test}}$.
We augment the calibration set by performing random temporal shuffling.
We also reverse the parallactic rotation vector so that (potentially unknown) real sources lose their temporal coherence with respect to the measurements.
In this setting, no synthetic sources are injected.

We draw a large number of observations ($100,000$) from $\mathcal{D}_{\text{calib}}$, apply our model to each of them, 
and concatenate all the outputs into a single vector $\widehat{\V x}_{\text{calib}}$ of size $n_{\text{calib}} = 100,000 \times 256 \times 256$ pixels. 
We can then compute the empirical pixel-wise cumulative distribution function (eCDF) of $\widehat{\V x}$ under $\mathcal{H}_0$ as:
\begin{equation}
  \forall \, \tau \in \mathbb{R}, \quad \widehat{F}_{\widehat{ \V x } | \mathcal{H}_0}(\tau) := \frac{1}{n_{\text{calib}}} 
  \sum_{i=0}^{n_{\text{calib}} -1} {\M 1}_{\widehat{\V x}_{\text{calib}, i} \leq \tau} \,.
\end{equation}
By the strong law of large numbers, it can be shown the eCDF converges to the true CDF:
\begin{equation}
  \forall \, \tau \in \mathbb{R}, 
  \quad \widehat{F}_{\widehat{ \V x  } | \mathcal{H}_0}(\tau) 
  \xrightarrow{n_{\text{calib}} \rightarrow +\infty}
  F_{\widehat{ \V x   } | \mathcal{H}_0}(\tau)\,,
\end{equation}
where $F_{\widehat{ \V x } | \mathcal{H}_0}$ denotes the true pixel-wise cumulative distribution function (CDF) of $\widehat{\V x}$ under $\mathcal{H}_0$:
\begin{equation}
  \forall \, \tau \in \mathbb{R}, \quad F_{\widehat{ \V x } | \mathcal{H}_0}
  :=  \mathbb{P}(\widehat{\V x} \leq \tau | \mathcal{H}_0)\,.
\end{equation}

In practice, we compute the eCDF with 500 threshold values (calibration points) for $\tau$.
Most of them are sampled in the tail of the distribution 
because we try to model the distribution of rare events under the null hypothesis.
We also discard extreme points as they are too noisy.

\begin{figure}
  \centering
  \includegraphics[width=.48\textwidth]{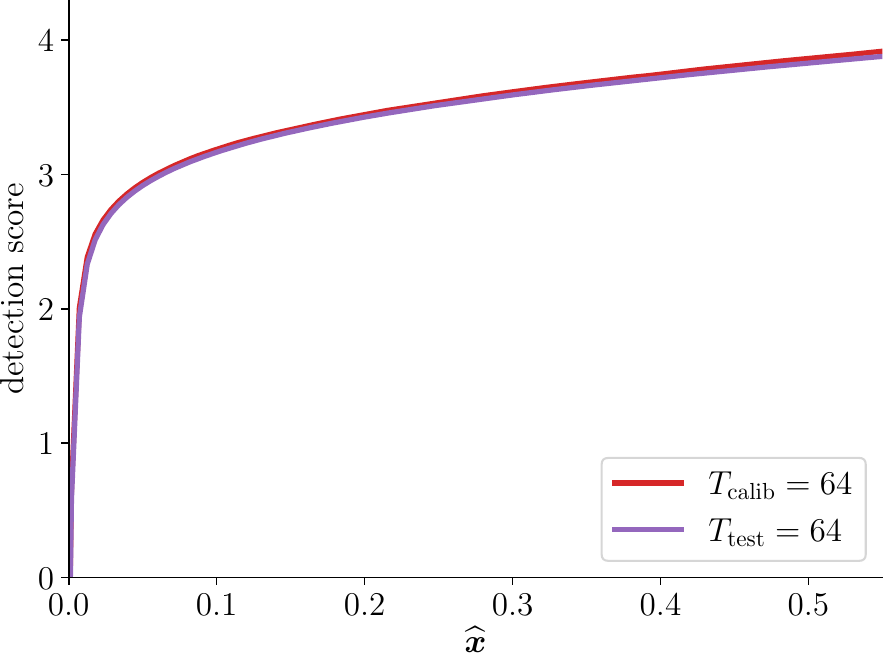}
    \caption{Comparison between calibration mapping $c$ performed on a calibration set and directly on a test set with observations of $T_{\text{calib}} = 64$ frames. 
}
    \label{fig:calib_test}
\end{figure}

\subsubsection{Hypothesis testing}
\label{sec:detection_score}

Once the CDF is approximated, we can apply hypothesis testing during inference to estimate the probability of false alarm (PFA) tied to a given threshold $\tau$ \citep{kay1993fundamentals}:
\begin{equation}
  \forall \, \tau \in \mathbb{R}, \quad \widehat{\text{PFA}}(\tau) = \widehat{\mathbb{P}}(\widehat{ \V x  } > \tau | \mathcal{H}_0) 
  = 1 - \widehat{F}_{\widehat{ \V x } | \mathcal{H}_0}(\tau)\,.
\end{equation}
However, the range of the calibration points is limited, because under the null hypothesis (i.e., no source injected), it is extremely rare to observe high values in the detection maps $\widehat{\V x}$.
In order to estimate the PFA for high values, we need to extrapolate $\widehat{\text{PFA}}$ to obtain the probability of extremely rare events under the null hypothesis.
Moreover, the extrapolated points should be in the $[0, 1]$ range.
To obtain a more practical representation, 
we map each threshold in the output space of our approach to its equivalent in terms of PFA for a standard normal distribution $ \mathcal{N}(0, 1)$ with zero mean and unit variance. 
This mapping can be accomplished using the CDF, as outlined by the \gls{probability_integral_transform_theorem}:
\begin{equation}
  c(\tau) = F^{-1}_{\mathcal{N}(0, 1)}\left( F_{\widehat{ \V x  } | \mathcal{H}_0}(\tau)\right)\,.
  \label{eq:calib_mapping}
\end{equation}
With this representation, calibration points can be extrapolated easily, as there is no constraint on the co-domain of the mapping.

\subsubsection{Experiments}
\label{subsubsec:calibration_experiments}

In Fig. \ref{fig:calib_n_frames}, we observe calibration curves corresponding to the mapping of Eq. \ref{eq:calib_mapping}, computed on the calibration set for different numbers of frames.
The calibration curves are similar when a sufficient number of frames are present ($T\geq 32$).
We adopt in the following the most conservative calibration, i.e. the one obtained for $T_\text{calib}=64$. 
This approach ensures that we can maintain control over the probability of false alarm (PFA) at or slightly above the designated detection threshold $\tau$, irrespective of the actual number of frames in the dataset. 

We check that the empirical distribution of $\widehat{\V x} | \mathcal{H}_0$ estimated on the 
calibration set matches with the test set.
To do so, we perform the same calibration procedure described in Sect. \ref{sec:calibration_set} on the test set, and compare both curves.
The results are displayed in Fig. \ref{fig:calib_test}. 
We observe that both curves remain close, even for a detection threshold at about 4.
The eCDF of the calibration set is thus a very good approximation of the eCDF of the test set.

\subsection{Implementation details}
\label{subsec:implementation_details}

The optimization of the network weights $\V \theta$ is performed with the deep learning library PyTorch \citep{NEURIPS2019_9015} on GPUs clusters equipped with NVIDIA systems with either Tesla V100 or GTX 1080 Ti cards.
The whitening of the data through the statistical modeling of spatial covariances embedded within the learnable modules of \thisalgo is also implemented on GPUs.
It prevents memory transfer between CPUs and GPUs and allows fast computation of matrix and element-wise products involved in Eqs.
(\ref{eq:shrinkage_convex_combination})-(\ref{eq:whitening}).
Under that framework, the training of our model typically takes a few hours with cubes of $64 \times 256 \times 256$ pixels.
The inference at evaluation time of the quantity $\widehat{\V x}$ from a new data cube $\V y$ typically takes a few seconds.
These specificities of the proposed algorithm offer practical advantages for its massive deployment on large-scale surveys.
We chose a batch size of 1, because of the high memory usage of data cubes.
To improve stability, we discard pixels in $\widehat{\V x}$ having received information from less than 8 frames, (i.e., corner pixels) in the loss function.

%% file: results.tex
\section{Results}
\label{sec:results}

\subsection{Datasets description and reduction strategies}
\label{subsec:datasets}

\begin{table*}
	\caption{Main observation parameters for the datasets from the VLT/SPHERE-IRDIS instrument considered in this paper. 
      Columns on the table correspond to: dataset ID we refer in this paper, target name, ESO survey ID, observation date, number $T_{\text{test}}$ of temporal frames used at test time, total apparent rotation $\Delta_{\phi}$ of the field of view, number NDIT of sub-integration exposures, individual exposure time DIT, average coherence time $\tau_0$, average seeing, and the first paper reporting analysis of the same data. Datasets \#1 to \#9 are used for massive simulations of synthetic sources, using either the native total parallactic rotation or four truncated versions obtained by keeping respectively the first $T_{\text{test}} \in \llbracket 64, 48, 32, 16 \rrbracket$ temporal frames of the sequence. The four corresponding total amounts of parallactic rotation are indicated inside parentheses in the field $\Delta_{\phi}$.
	}
	\label{tab:dataset_logs}
	\centering
	\begin{tabular}{ccccccccccc}
		\toprule
		\#ID              & Target     & ESO ID         & Obs. date  & $T_{\text{test}}$ & $\Delta_{\phi}$ total ; (truncated) & NDIT & DIT & $\tau_0$ & Seeing & Related paper                       \\
		                  &            &                &            &                   & (°)                                 &      & (s) & (ms)     & ('')   &                                     \\
		\midrule
		\multicolumn{11}{c}{\textit{Datasets used for experiments relying on massive injections of synthetic sources, see Sects. \ref{subsec:detection_synthetic_sources} and \ref{subsec:ablation}}} \\
		\midrule
		\#1               & HD 159911  & 097.C-0865(A)  & 2016-04-16 & 68                & 54.6 (53.6, 46.7, 36.8, 16.7)       & 10   & 64  & 4.8      & 0.36   & \cite{langlois2021sphere}           \\
		\#2               & HIP 27288  & 198.C-0209(E)  & 2017-02-11 & 210               & 82.5 (25.5, 19.0, 14.9, 11.1)       & 18   & 16  & 5.2      & 0.47   & \cite{langlois2021sphere}           \\
		\#3               & HIP 50191  & 1100.C-0481(D) & 2018-02-27 & 116               & 52.3 (24.4, 17.4, 10.1, 4.5)        & 20   & 32  & 7.9      & 0.42   & \cite{langlois2021sphere}           \\
		\#4               & HIP 112312 & 095.C-0298(D)  & 2015-10-01 & 93                & 54.7 (19.1, 11.9, 6.7, 2.9)         & 8    & 32  & 2.3      & 0.36   & \cite{langlois2021sphere}           \\
		\#5               & HIP 37288  & 095.C-0298(D)  & 2017-02-13 & 69                & 38.4 (36.2, 29.6, 20.0, 10.5)       & 5    & 64  & 10.5     & 0.58   & \cite{langlois2021sphere}           \\
		\#6               & HIP 113283 & 095.C-0298(C)  & 2015-06-27 & 319               & 100.7 (22.8, 18.5, 6.3, 2.9)        & 32   & 8   & 2.5      & 0.38   & \cite{langlois2021sphere}           \\
		\#7               & HIP 107350 & 095.C-0298(D)  & 2015-09-30 & 182               & 25.7 (10.8, 8.5, 6.9, 4.5)          & 16   & 16  & 3.4      & 0.44   & \cite{langlois2021sphere}           \\
		\#8               & HIP 98495  & 095.C-0298(B)  & 2015-06-01 & 228               & 23.2 (5.8, 4.0, 2.6, 1.2)           & 20   & 16  & 1.2      & 0.75   & \cite{langlois2021sphere}           \\
		\#9               & HIP 57632  & 095.C-0298(B)  & 2015-05-31 & 843               & 29.7 (1.7, 1.3, 0.9, 0.4)           & 64   & 4   & 1.3      & 1.32   & \cite{langlois2021sphere}           \\
		\midrule
		\multicolumn{11}{c}{\textit{Datasets used for experiments on known real sources, see Sect. \ref{subsec:detection_real_sources}}}                                                              \\
		\midrule
		\#10              & HIP 65426  & 097.C-0865(B)  & 2016-05-31 & 43                & 34.2                                & 4    & 64  & 2.1      & 0.67   & \cite{chauvin2017discovery}         \\
		\#11              & HIP 65426  & 198.C-0209(D)  & 2017-02-06 & 54                & 44.3                                & 8    & 64  & 7.2      & 0.39   & \cite{chauvin2017discovery}         \\
		\#12              & HIP 88399  & 095.C-0298(A)  & 2015-05-10 & 46                & 34.3                                & 4    & 64  & 1.2      & 1.05   & \cite{langlois2021sphere}           \\
		\#13              & HIP 88399  & 097.C-0865(A)  & 2016-04-16 & 54                & 37.3                                & 5    & 64  & 2.0      & 1.45   & \cite{langlois2021sphere}           \\
		\#14              & HIP 88399  & 1100.C-0481(F) & 2018-04-11 & 40                & 31.9                                & 10   & 96  & 5.5      & 0.74   & \cite{langlois2021sphere}           \\
		\#15              & HR 8799    & 198.C-0209(B)  & 2016-11-17 & 38                & 16.8                                & 16   & 32  & 4.24     & 0.93   & \cite{langlois2021sphere}           \\
		\#16              & HR 8799    & 198.C-0209(J)  & 2017-06-14 & 84                & 19.3                                & 48   & 24  & 7.1      & 0.65   & \cite{langlois2021sphere}           \\
		\#17              & HR 8799    & 1100.C-0481(H) & 2018-06-18 & 38                & 34.4                                & 8    & 96  & 7.8      & 0.67   & \cite{langlois2021sphere}           \\
		\#18$^{\text{a}}$ & HD 95086   & 095.C-0298(A)  & 2015-05-05 & 38                & 22.4                                & 4    & 64  & 2.3      & 0.74   & \cite{dallant2023multi}             \\
		\bottomrule
	\end{tabular}
	\begin{tablenotes} \item  \hspace{-5mm}
		\textbf{Notes.}
		All the observations were performed with the apodized Lyot coronagraph \citep{carbillet2011apodized} of the VLT/SPHERE-IRDIS instrument in the H2-H3 spectral band.
		$^{\text{a}}$HD 95086 was observed both in the H2-H3 and K1-K2 spectral bands during the 2015-05-05 night. As the exoplanet HD 95086 b can be easily detected from the H2-H3 observations, this dataset was considered by multiple works, see for instance \citep{chauvin2018investigating, desgrange2022depth}. However, as the exoplanet HD 95086 b was not detectable in the H2-H3 observations (2015-05-05), dataset \#18 was used for the first time in \citep{dallant2023multi} for multi-epochs combination of PACO reductions, but HD 95086 b remained undetectable based on the analysis of the single epoch of 2015-05-05, as also reported in Sect. \ref{subsec:detection_real_sources}.
	\end{tablenotes}
  \label{table:observations}
\end{table*}

For our comparisons, we selected 18 datasets obtained under various observing conditions with the SPHERE-IRDIS instrument.

In Sect. \ref{subsec:detection_synthetic_sources}, we evaluate quantitatively the performance of the proposed approach compared to the PACO algorithm on nine datasets (numbered \#1 to \#9)  obtained by the observations of stars where no (candidate) point-like sources are reported in the high-contrast literature within the considered field of view. 
We thus can use these datasets to perform massive simulations of synthetic sources with little chance of being biased by real astrophysical signals. 
As an additional precaution for these experiments, we consider simulated parallactic angles differing from the natural ones, so that any real astrophysical signal (even undetectable with existing post-processing algorithms) will not co-add constructively when de-rotating and stacking the temporal frames. 
Simulating parallactic angles also allows us to study the performance of the proposed approach when facing different amounts of total rotation induced by ADI. In Sect. \ref{subsec:ablation} we perform a model \gls{ablation_study} on the proposed approach using these nine datasets. 
We study in Sect. \ref{subsec:obs_cond} the influence of the experienced observing conditions on the learning strategy of the model. 

In Sect.~\ref{subsec:detection_real_sources}, we consider nine additional datasets (numbered \#10 to \#18) obtained from observations of four stars (some observed multiple times). 
These stars either host emblematic exoplanet(s) or are known to have faint background sources within the projected field of view of observations.
These datasets are used to qualitatively analyze the benefits of the proposed algorithm in detecting faint sources compared to the PACO and deep PACO algorithms.
For this analysis, the four selected stars are the following:\\
-- HIP 65426, a A8III type star of the Carina constellation, hosting an exoplanet (HIP 65426 b) discovered by direct imaging with the SPHERE instrument \citep{chauvin2017discovery}.\\
-- HIP 88399, a F6V type star of the Vela constellation, without known gravitationally bounded exoplanet. Six faint background sources fall however within the SPHERE-IRDIS field of view \citep{langlois2021sphere}. Among them, one (denoted \textit{bkg} in the following) falls within the field of view considered in this work.\\
-- HR 8799, a A5V type star of the Pegasus constellation, hosting four exoplanets discovered by direct imaging \citep{marois2008direct, marois2010images}. Depending on the epoch, three of them (HR 8799 c, d, and e) or all four known exoplanets (HR 8799 b, c, d, and e) are within the field of view considered in this work.\\
-- HD 95086, a A8III type star of the Carina constellation, hosting an exoplanet (HD 95086 b) discovered by direct imaging with the VLT/NaCo instrument \citep{rameau2013discovery,rameau2013confirmation}.

Table \ref{tab:dataset_logs} summarizes the main observation parameters associated with these 18 datasets.

\subsection{Evaluation metrics}
\label{subsec:evaluation_metrics}

We evaluate in our experiments the ability of the model to detect faint point-like sources while simultaneously avoiding false alarms. 
In other words, the model should reach a trade-off between recall and precision. 
To measure this trade-off, we count the number of true positives (TP, i.e., true detections), false positives (FP, i.e., false alarms) and false negatives (FN, i.e., missed detections) from either a reconstructed flux distribution map $\widehat{\V x}$ (output of the proposed approach) thresholded at $\tau \in \left[0 ; \text{max}_n ( \widehat{\V x}_n ) \right]$, a S/N map (output of PACO) thresholded at  $\tau \in \left[\text{min}_n ( \text{S/N}_n ) ; \text{max}_n ( \text{S/N}_n ) \right]$, or a pseudo-probability map (output of deep PACO) thresholded at $\tau \in \left[0 ; 1 \right]$. 
Based on standard practices in direct imaging (see e.g., \cite{flasseur2018exoplanet,gonzalez2018supervised,cantalloube2020exoplanet}), 
a blob in the detection map is classified as a true positive if it lies within one resolution element from one of the ground truth locations. 
The size of one resolution element, defined as the \textit{full width at half-maximum} (FWHM) of the off-axis PSF, corresponds to the expected spatial extent of an exoplanetary signature.
From TP, FP, and FN, we compute the false discovery rate (FDR) and the true positive rate (TPR):
\begin{equation}
	\text{FDR} = \frac{\text{FP}}{\text{FP} + \text{TP}}\in [0;1]\,,\,\,\,\,\, \text{TPR} = \frac{\text{TP}}{\text{TP} + \text{FN}} \in [0;1]\,.
	\label{eq:detection_metrics}
\end{equation}
We build so-called \textit{receiver operating curves} (ROCs) by reporting the TPR as a function of the FDR by varying the detection threshold $\tau$. 
This metric is used routinely in the signal-processing community (see e.g. \cite{kay1993fundamentals}) to evaluate the performance of a detector, and its relevance in the context of exoplanet imaging has been illustrated by several works, see e.g. \cite{gonzalez2016low,gonzalez2018supervised,flasseur2018exoplanet,dahlqvist2020regime,cantalloube2020exoplanet,daglayan2022likelihood}.
Finally, we compute the area under the resulting curve (AUC, best when close to 1). 
It corresponds to an aggregated measurement of the precision-recall capability of the model, averaged over all possible FDR (i.e., detection confidence levels). 
As a high confidence level is required to claim a new detection in direct imaging (typically, with a S/N of detection higher than 5), we will also report the TPR for $\text{FDR}=0$ and we will denote this quantity as $\text{TPR}_0$ in the following. 
It corresponds to the fraction of true sources actually detected by the model with a detection threshold selected to experience no false alarm on the whole field of view.

\subsection{Detection of synthetic sources}
\label{subsec:detection_synthetic_sources}

\begin{table}
	\centering
	\caption{Flux in contrast units (minimum $\alpha_{\text{min}}$, maximum $\alpha_{\text{max}}$, average $\alpha_{\text{mean}}$) for the synthetic sources considered in the experiments reported in Figs. \ref{fig:23deg}-\ref{fig:2deg}. 
  The name of the target star and the total parallactic rotation $\Delta_\phi$ associated with the corresponding datasets are also recalled.}
	\begin{tabular}{ccccc}
		\toprule
		Target\;   & $\Delta_\phi (\degree)$ & $\alpha_{\text{min}}$\; & $\alpha_{\text{max}}$\; & $\alpha_{\text{mean}}$ \\
		\midrule
		HIP 113283 & 22.8                    & $3.14\times 10^{-7}$     & $2.48\times 10^{-5}$     & $2.48\times 10^{-5}$    \\
		HIP 107350 & 10.8                    & $3.48\times 10^{-7}$     & $1.06\times 10^{-5}$     & $2.15\times 10^{-6}$    \\
		HIP 98495  & 5.8                     & $1.37\times 10^{-6}$     & $2.81\times 10^{-4}$     & $2.09\times 10^{-5}$    \\
		HIP 57632  & 1.7                     & $3.07\times 10^{-6}$     & $2.22\times 10^{-4}$     & $3.80\times 10^{-5}$    \\
		\bottomrule
	\end{tabular}
	\label{tab:flux_infos_parallactic_expes}
\end{table}

\begin{samepage}
	\begin{figure*}
		\centering
		\includegraphics[width=\textwidth]{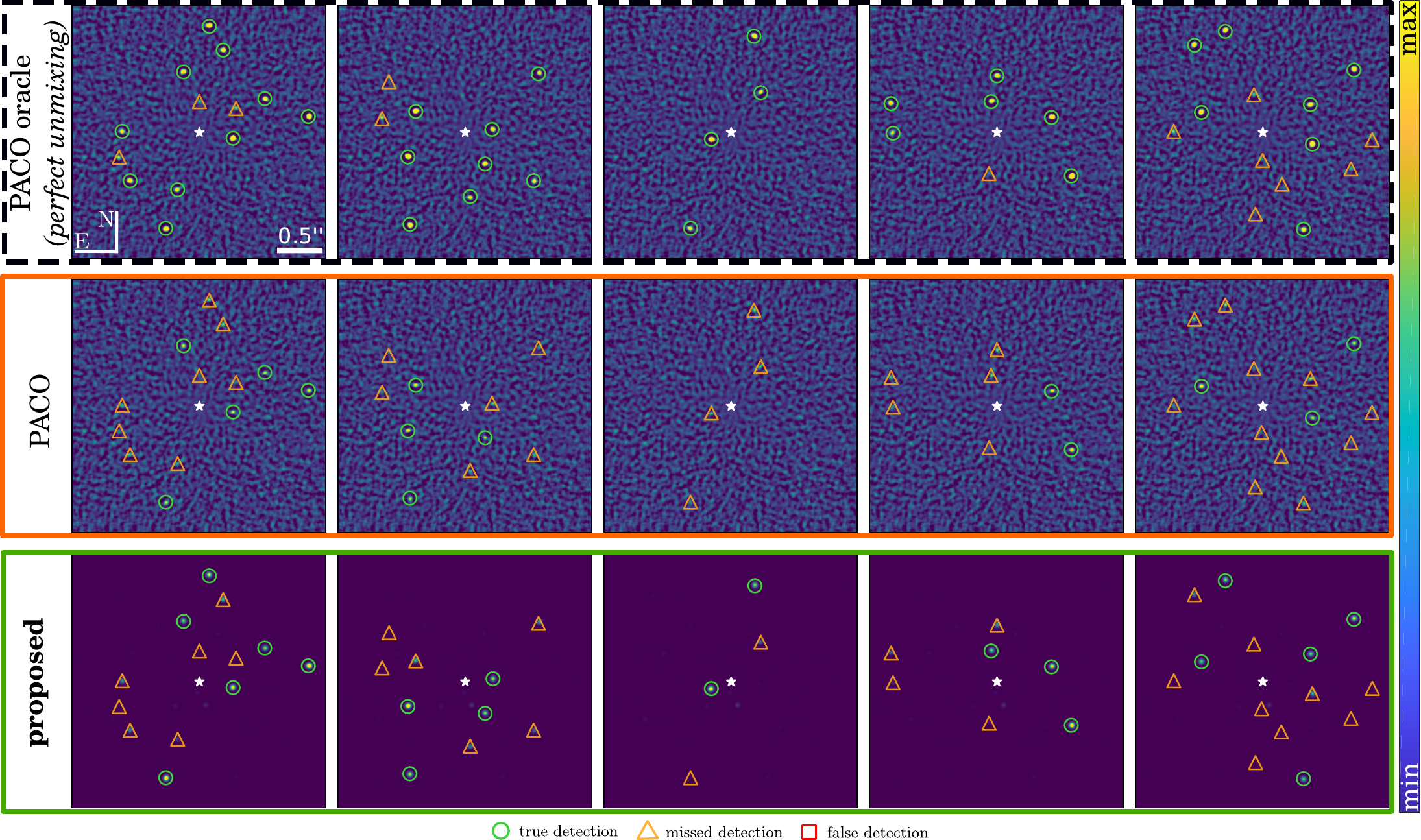}
		\caption{Detection maps for five configurations of simulated point-like sources whose locations are pointed out by circles. Reconstruction results $\widehat{\V x}$ obtained with the proposed approach (bottom) are compared to S/N maps obtained with PACO (middle) and with the oracle version of PACO (top) assuming a perfect unmixing (not achievable in practice) between the objects of interest and the nuisance component for the computation of the nuisance parameters. Dataset: HIP 113283 (2015-06-27, \#6), total parallactic rotation $\Delta_{\phi}=22.8\degree$ for $T_{\text{test}}=64$ temporal frames kept, see Sect. \ref{subsec:datasets} for the detailed observation parameters. 
    We used a detection threshold of 5 for all methods.}
		\label{fig:23deg}

		\bigskip

		\centering
		\includegraphics[width=\textwidth]{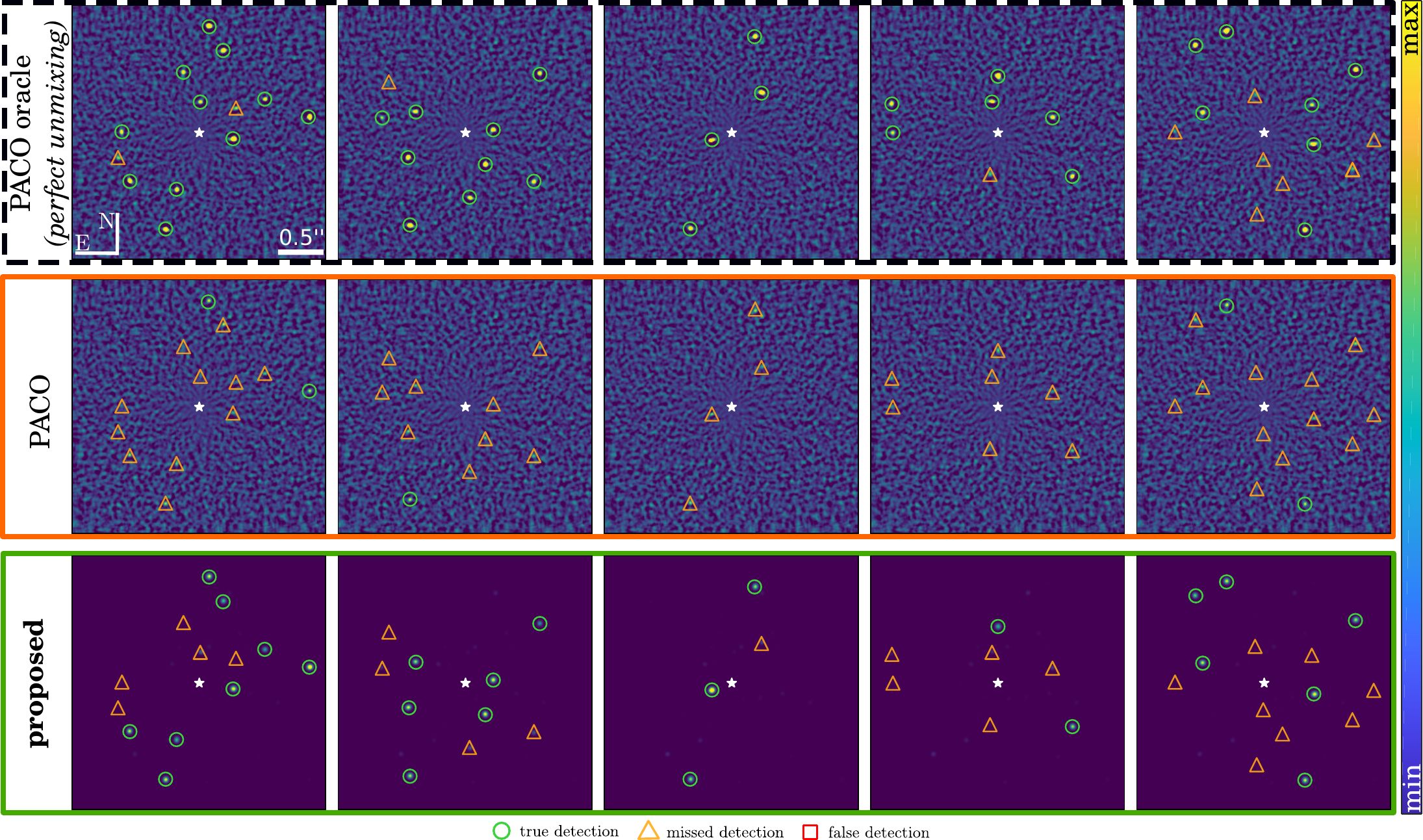}
		\caption{Same as Fig. \ref{fig:23deg} for $\Delta_{\phi}=10.8\degree$. Dataset: HIP 107350 (2015-09-30, \#7), see Sect. \ref{subsec:datasets} for the observation parameters. }
		\label{fig:10deg}
	\end{figure*}
\end{samepage}

\begin{samepage}
	\begin{figure*}
		\centering
		\includegraphics[width=\textwidth]{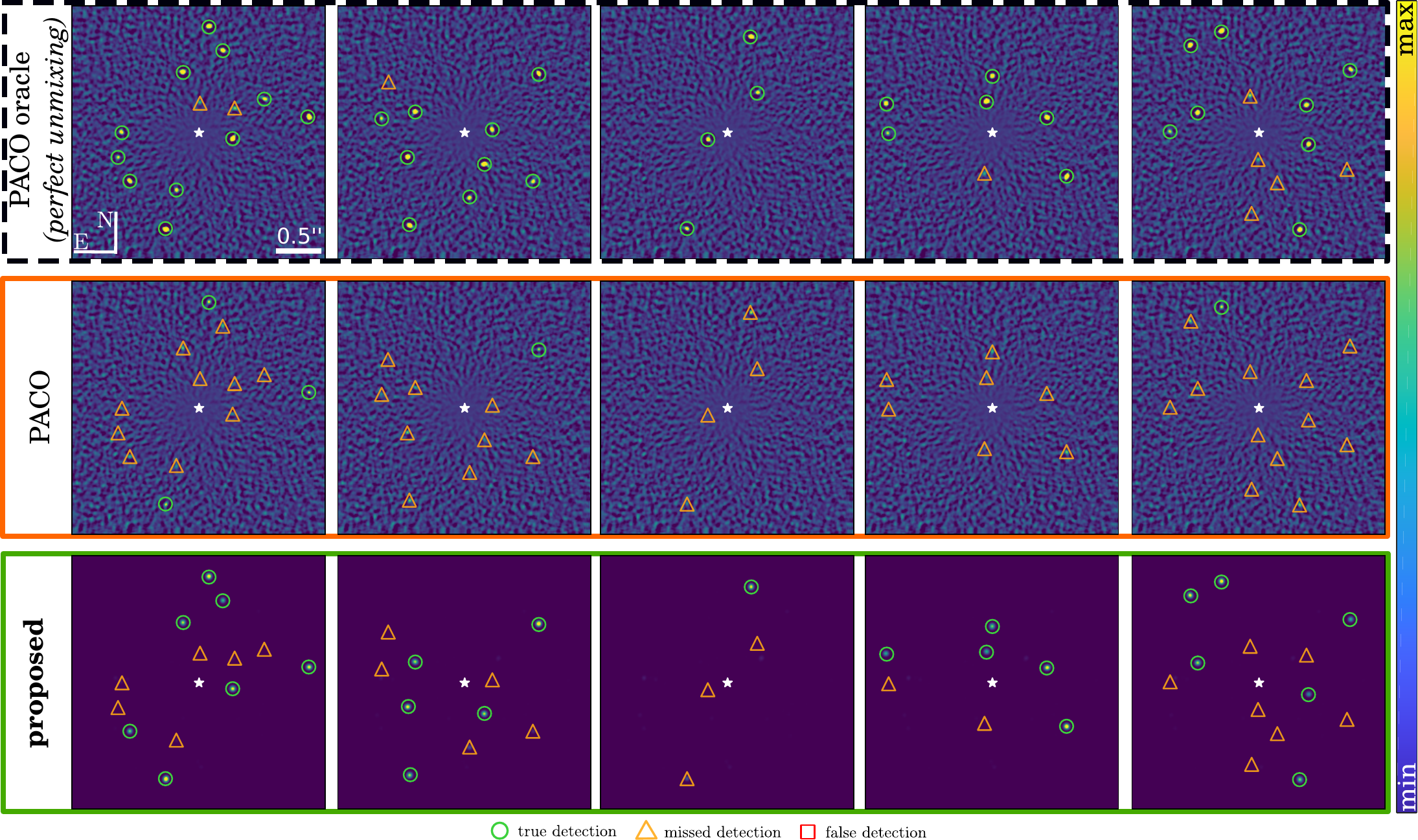}
		\caption{Same as Fig. \ref{fig:23deg} for $\Delta_{\phi}=5.8\degree$. Dataset: HIP 98495 (2015-06-01, \#8), see Sect. \ref{subsec:datasets} for the observation parameters.}
		\label{fig:6deg}

		\bigskip

		\centering
		\includegraphics[width=\textwidth]{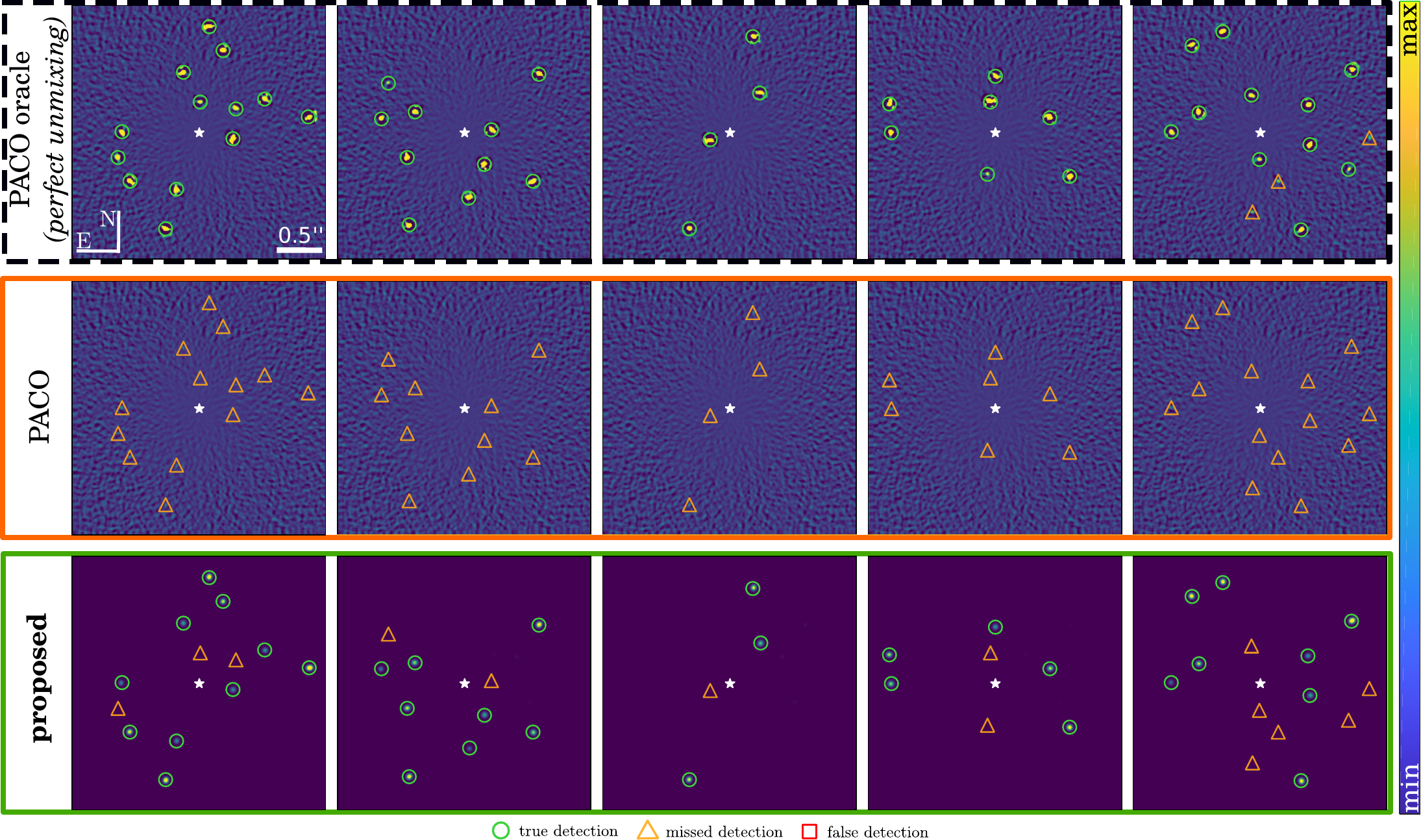}
		\caption{Same as Fig. \ref{fig:23deg} for $\Delta_{\phi}=1.7\degree$. Dataset: HIP 57632 (2015-05-31, \#9), see Sect. \ref{subsec:datasets} for the observation parameters.}
		\label{fig:2deg}
	\end{figure*}
\end{samepage}

We start by evaluating qualitatively the performance of the proposed \thisalgo algorithm.
For that purpose, we resort to the massive simulation and injection of synthetic sources within nine real VLT/SPHERE-IRDIS datasets (\#1 to \#9, see Sect. \ref{subsec:datasets}). 
For the purpose of illustration, four of these datasets (\#6 to \#9) are successively considered in Figs. \ref{fig:23deg} to \ref{fig:2deg} by keeping the first $T_{\text{test}} = 64$ temporal frames from the complete sequences of observation.
Five configurations of synthetic sources are drawn randomly for each dataset to ensure diversity.
The amplitude $\Delta_\phi$ of the parallactic rotation associated with the considered datasets decreases from Fig. \ref{fig:23deg} ($\Delta_{\phi}=22.8\degree$) to Fig. \ref{fig:2deg} ($\Delta_{\phi}=1.7\degree$), such that the angular diversity brought by ADI is also reduced.
Table \ref{tab:flux_infos_parallactic_expes} summarizes the relative flux information about synthetic sources simulated for these experiments.

In each case, we compare the estimated signal $\widehat{\V x}$ reconstructed with \thisalgo to the S/N map produced by the PACO algorithm.
We also compare \thisalgo to the so-called \textit{oracle} version of PACO, where the nuisance parameters are estimated without the presence of a source.
This optimal regime is not achievable in practice as it corresponds to a situation where the bias (i.e., the self-subtraction) due to the lack of angular diversity at short angular separations is always null, whatever the total amount $\Delta_{\phi}$ of parallactic rotation.
Oracle PACO is thus only limited by the fidelity of the multi-variate Gaussian model to the actual observations, and not anymore by the estimation of the underlying parameters (mean and covariances).

The qualitative results illustrated by Figs. \ref{fig:23deg} to \ref{fig:2deg} suggest that the proposed approach is able to recover synthetic sources competitively with the PACO algorithm on standard observations with typically more than 20 degrees of total parallactic rotation.
In this setting, some sources remain non-detectable by both PACO and the proposed approach while they are clearly detectable with the oracle mode of PACO (even though this oracle performance can not be achieved in practice).
This indicates that a residual bias due to the lack of angular diversity remains in both approaches. 
However, this residual bias is significantly reduced by the proposed approach leveraging multiple observations to build the model of the nuisance.
As an illustration, decreasing the total parallactic rotation $\Delta_{\phi}$ until two degrees for sixty-four frames leads to barely detectable point-like sources with PACO, while the proposed approach displays source signals significantly closer than the oracle capability.

These results exemplify the benefits of our approach to learning useful information from multiple observations in order to limit the self-subtraction phenomenon occurring at short angular separations with any observation-dependent algorithm.

\begin{figure*}
	\centering
	\includegraphics[width=\textwidth]{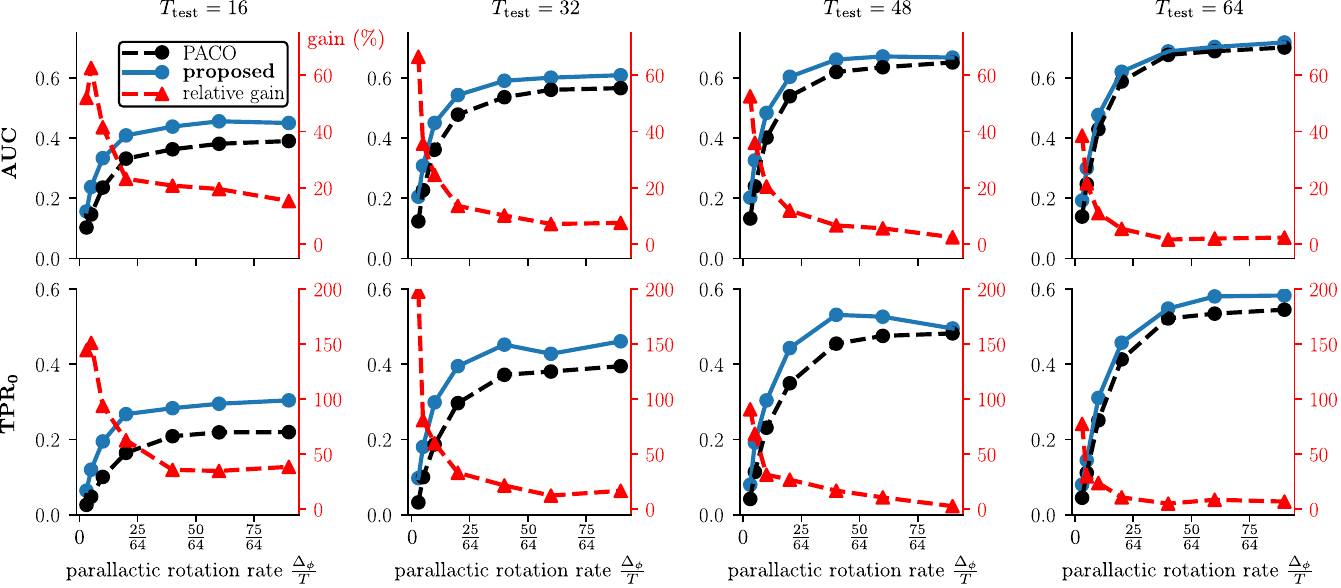}
  \caption{Evolution of the AUC (first line) and $\text{TPR}_0$ (second line) metrics as the function of the total amount of parallactic rotation $\Delta_{\phi}$. From left to right, different numbers $T_\text{test}$ of frames are considered within the datasets at testing time. For each graph, the relative gain (in \%) brought by \thisalgo with respect to PACO is reported with the red line. Results are averaged over nine SPHERE-IRDIS datasets \#1 to \#9, see Sect \ref{subsec:datasets} for their respective observation parameters.}
	\label{fig:avg_auc_fdr_tpr_main_fullfig}
\end{figure*}

\medskip

\noindent After this qualitative study, we now aim at grounding quantitatively the performance of the proposed approach on synthetic sources.
We evaluate our method under multiple parallactic rotation rates $\Delta_\phi / T_\text{test}$, and a total number of frames $T_\text{test}$.
We chose $T_\text{test} \in \llbracket 16, 32, 48, 64 \rrbracket$,
and the rotation rates such that $\Delta_\phi \in \llbracket 3\degree, 5\degree, 10\degree, 20\degree, 40\degree, 60\degree, 90\degree \rrbracket$ when $T_\text{test}=64$.

Figure \ref{fig:avg_auc_fdr_tpr_main_fullfig} reports the AUC and $\text{TPR}_\text{0}$ averaged over the nine datasets \#1 to \#9 of SPHERE-IRDIS (see Sect.\ref{subsec:datasets}) as a function of the total parallactic rotation $\Delta_{\phi}$. Appendix \ref{app:detailed_rocs_results} complements Fig. \ref{fig:avg_auc_fdr_tpr_main_fullfig} with the detailed results obtained on each of the nine datasets considered in this study.
The \thisalgo model is trained on a dataset with $T_{\text{clip}}=64$ frames, which corresponds to the best possible setting for our approach given computational constraints, see Sect. \ref{subsubsec:clips}
 for the clip extraction procedure.

These results emphasize that the proposed approach outperforms the PACO algorithm in terms of precision-recall trade-off, both on the averaged AUC and $\text{TPR}_\text{0}$ criteria, whatever the values of the varying parameters $\Delta_\phi$ and $T_{\text{test}}$.
Interestingly, the relative gain between the two methods is higher for small parallactic rotation rates $\frac{\Delta_{\phi}}{T}$  and also for datasets having a small number $T_{\text{test}}$ of temporal frames, i.e.
in the typical regime where the detection is known to be the most challenging for observation-dependent algorithms like PACO.
As an example, for $\frac{\Delta_{\phi}}{T} = \frac{5}{64}$, the mean gain on the AUC (respectively, $\text{TPR}_\text{0}$) metric is between 50\% and 75\% (respectively, 80\% and 200\%), depending on the number $T_{\text{test}}$ of temporal frames.
These quantitative results confirm the qualitative observations derived from the detection maps presented in Figs. \ref{fig:23deg}-\ref{fig:2deg}.

The improvement achieved by the proposed approach, particularly important when the parallactic rotation is the most limited, is attributed to \thisalgo's ability to partially compensate for the lack of angular diversity by leveraging a model derived from multiple observations.
Similarly, the gain in detection performances observed for a smaller number $T_{\text{test}}$ of temporal frames is also a consequence of angular diversity limitations affecting the performance of any observation-dependent model.
This phenomenon is also the result of the specific shrinkage of the data covariances implemented in PACO, as detailed in Eqs.(\ref{eq:shrinkage_convex_combination})-(\ref{eq:whitening}).
Indeed, the shrinkage amount $\widehat{\rho}$ is higher when the number of temporal frames is small, thus mitigating the large variance (noise) affecting the estimated covariances $\widehat{\M C}$.
This shrinkage ensures the invertibility of the covariances, but as a side-effect, the detection sensitivity is downgraded because the covariances of the nuisance are partly ignored (i.e., the factor $\widehat{\rho}$ reaches a bias-variance trade-off).
The proposed approach does not suffer from this limitation because the modeling of the covariances is performed on the learnable features (of fixed size) of the speckles-aligned module rather than on the data themselves.

\medskip

\begin{figure*}
	\centering
	\includegraphics[width=.9\textwidth]{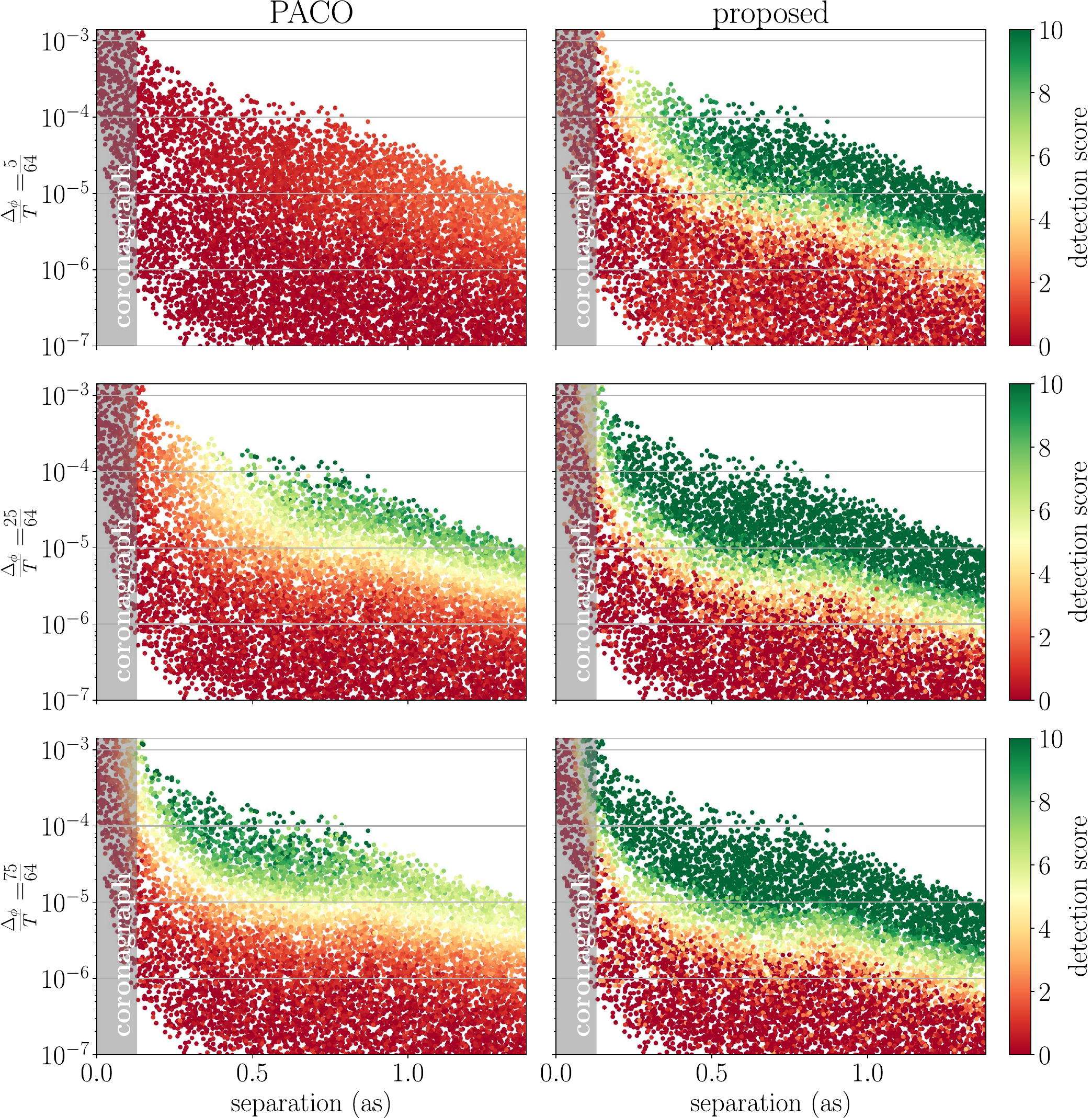}
	\caption{Flux (in contrast unit) as the function of the angular separation. Each point represents a simulated exoplanet injected within the data. Corresponding detection scores (in S/N unit) are color-coded, both for PACO (left) and the proposed \thisalgo algorithm. The gray area represents the inner working angle of the coronagraphic mask. Dataset: HIP 50191 (2018-02-28, \#3), see Sect. \ref{subsec:datasets} for the observation parameters. The number of temporal frames kept at test time is $T_{\text{test}} = 16$.}
	\label{fig:contrast_hd88_T16}
\end{figure*}

\begin{figure*}
	\centering
	\includegraphics[width=.9\textwidth]{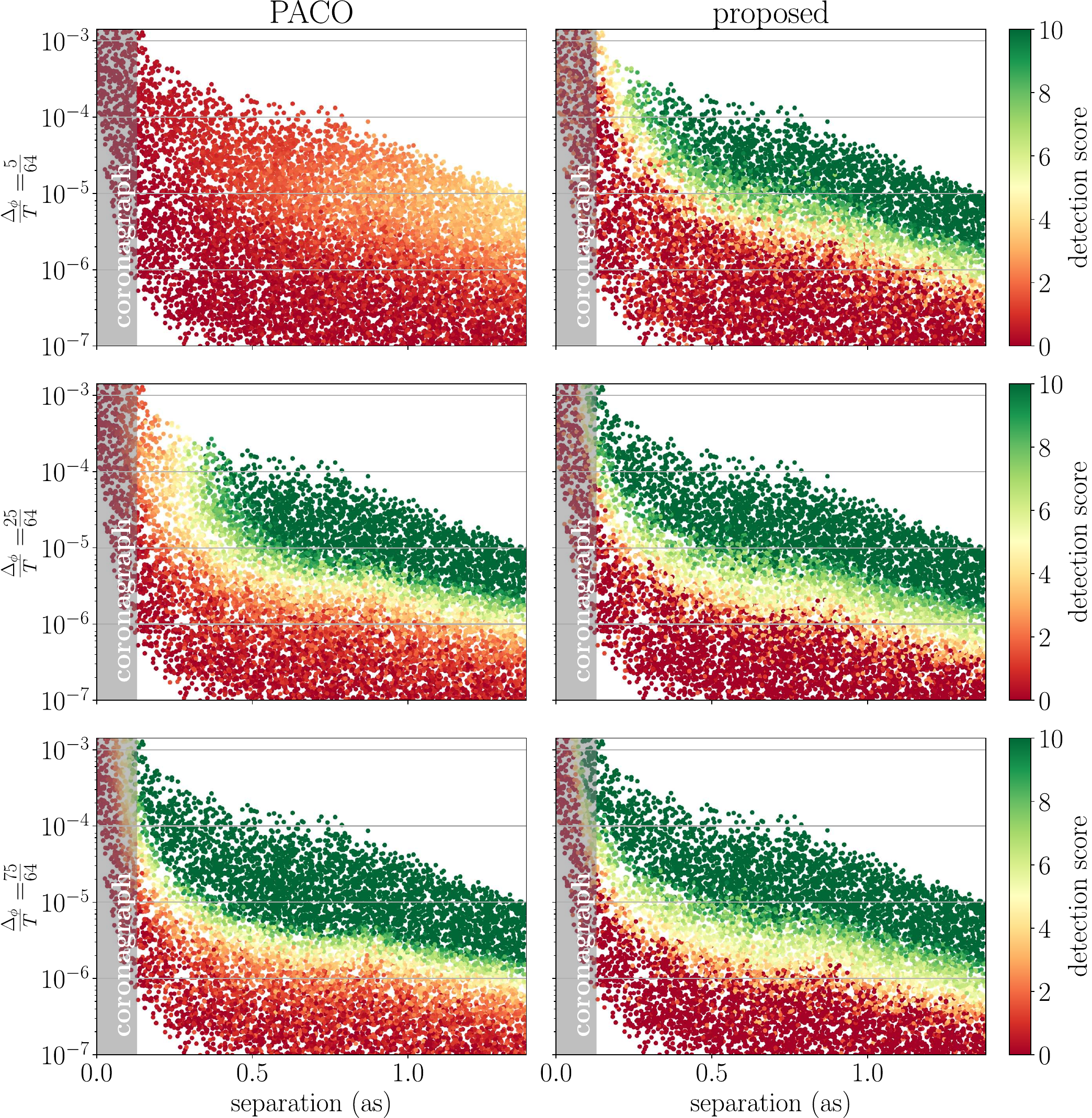}
	\caption{Same as Fig. \ref{fig:contrast_hd88_T16}, but for $T_{\text{test}} = 48$ temporal frames at test time.}
	\label{fig:contrast_hd88_T48}
\end{figure*}

\begin{figure*}
	\centering
	\includegraphics[width=.9\textwidth]{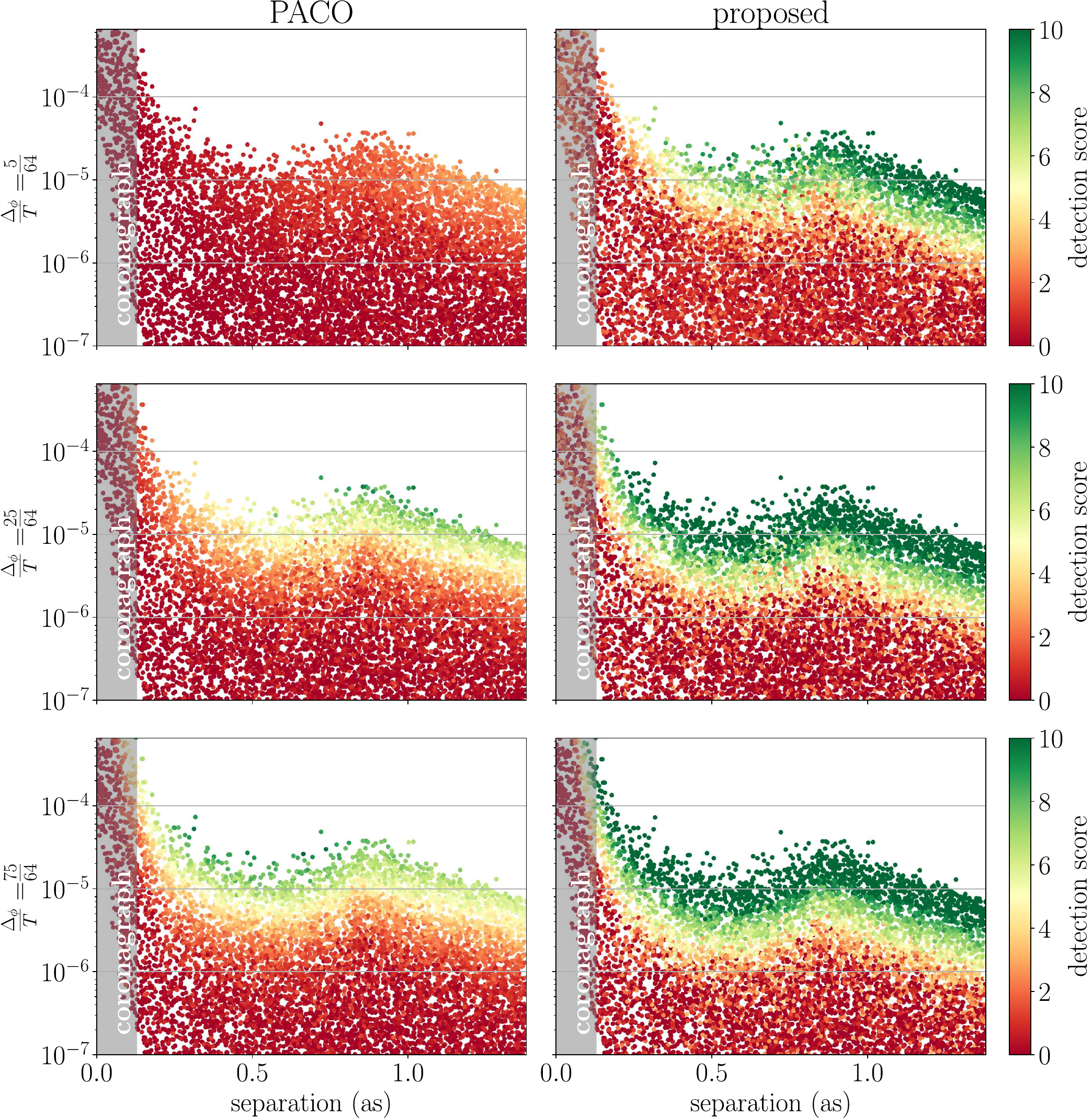}
	\caption{Flux (in contrast unit) as the function of the angular separation. Each point represents a simulated exoplanet injected within the data. Corresponding detection scores (in S/N unit) are color-coded, both for PACO (left) and the proposed \thisalgo algorithm. The gray area represents the inner working angle of the coronagraphic mask. Dataset: HIP 57632 (2015-05-31, \#9), see Sect. \ref{subsec:datasets} for the observation parameters. The number of temporal frames kept at test time is $T_{\text{test}} = 16$.}
	\label{fig:contrast_hd10_T16}
\end{figure*}

\begin{figure*}
	\centering
	\includegraphics[width=.9\textwidth]{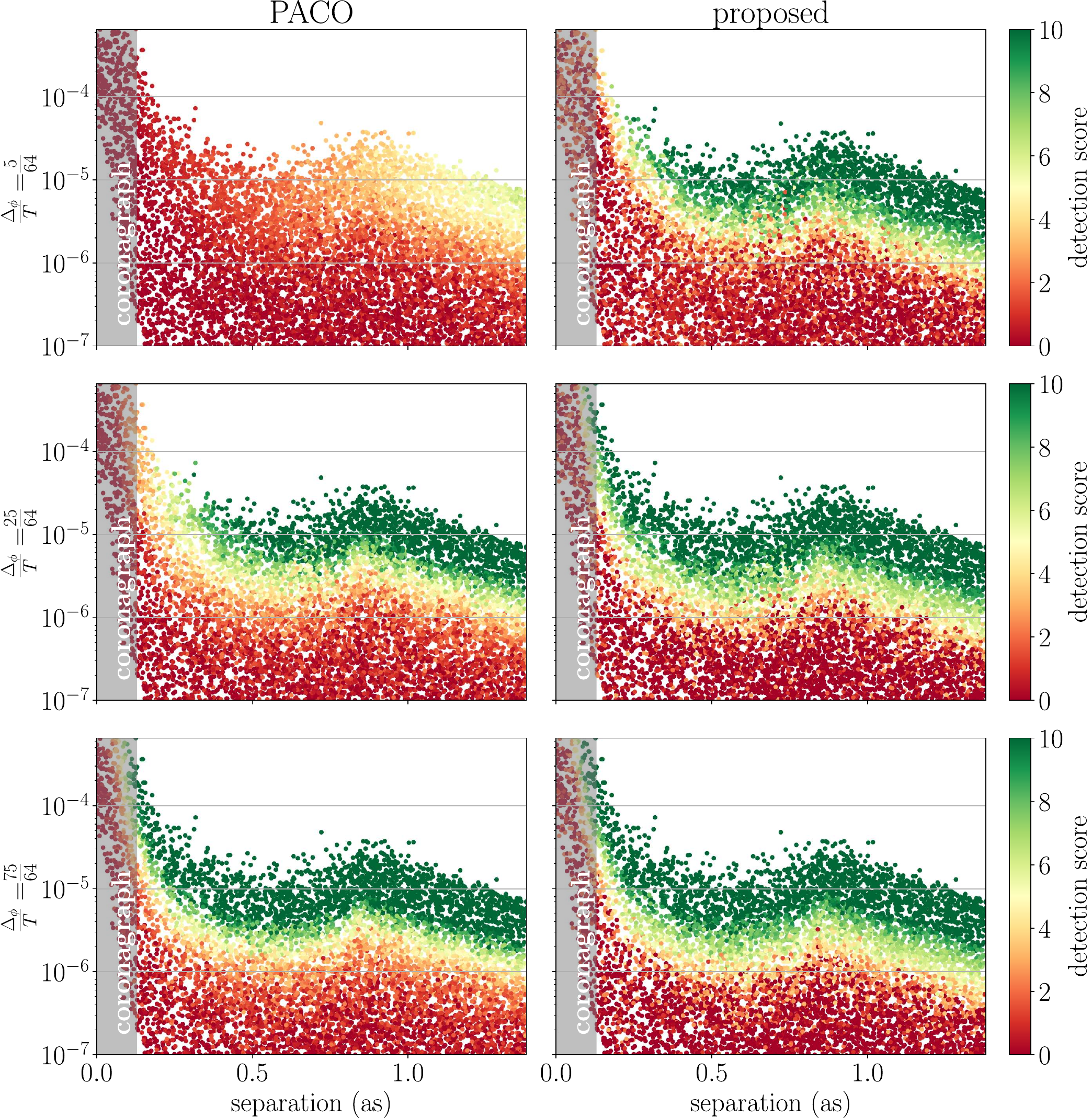}
	\caption{Same as Fig. \ref{fig:contrast_hd10_T48}, but for $T_{\text{test}} = 48$ temporal frames at test time.}
	\label{fig:contrast_hd10_T48}
\end{figure*}

\noindent We now derive empirical contrast curves for the proposed approach comparatively to the PACO algorithm.
For that purpose, we represent in the parameter space \textit{source flux} (in contrast unit) versus \textit{angular separation}, and the calibrated detection score (i.e., with PFA statistically controlled) for each simulated synthetic source.
We performed these experiments for each of the nine datasets (\#1 to \#9) described in Sect.\ref{subsec:datasets},
by varying both the number of individual frames in the ADI sequence ($T_{\text{test}} \in \llbracket 16, 48\rrbracket$ frames) and the parallactic rotation rate ($\frac{\Delta_{\phi}}{T} \in \llbracket \frac{5}{64}, \frac{25}{64}, \frac{75}{64} \rrbracket$).
Typical results obtained from two of the nine datasets are reported in Figs.
\ref{fig:contrast_hd88_T16}, \ref{fig:contrast_hd88_T48}, \ref{fig:contrast_hd10_T16}, and \ref{fig:contrast_hd10_T48}.
When both the number of temporal frames is limited (typically, $T_{\text{test}} = 16$, and the total amount of parallactic rotation is small (typically, $\frac{\Delta_{\phi}}{T} = \frac{5}{64}$), the proposed approach improves the achievable contrast by a factor eight to ten for all tested angular separations (i.e., within 0.13'' to 1.5'').
For moderate amounts of parallactic rotation (typically, $\frac{\Delta_{\phi}}{T} = \frac{25}{64}$), a similar gain is only observed typically below 0.35''.
At larger angular separations, PACO and the proposed approach lead to similar levels of contrast.
For larger amounts of parallactic rotation, PACO and the proposed approach lead on average to comparable results because we are in a regime where the bias introduced by self-subtraction is negligible given the diversity induced by ADI.
In this configuration, the best performing algorithm depends on the datasets, and the slight advantage or disadvantage brought by \thisalgo in terms of achievable contrast is directly related to the accuracy of its calibration procedure, which is global over the whole field of view (unlike PACO, see Sect. \ref{subsec:calibration}).
When the number of temporal frames increases, similar overall conclusions can be drawn with a slightly decreased typical gain brought by \thisalgo (typically, a gain in achievable contrast with respect to PACO up to a factor five to eight).

These general conclusions (clear gain brought by the proposed approach at small angular separations, mostly driven by the amplitude of the parallactic rotation) are also consistent with ROCs results presented in Fig. \ref{fig:avg_auc_fdr_tpr_main_fullfig}.

\subsection{Model ablation analysis}
\label{subsec:ablation}

\begin{table*}
	\centering
	\begin{tabular}{lcccc}
		\toprule
		method  & $T_{\textrm{test}}=16$ & $T_{\textrm{test}}=32$ & $T_{\textrm{test}}=48$ & $T_{\textrm{test}}=64$ \\
		\midrule
		PACO            & 0.251                  & 0.363                  & 0.410                  & 0.441                  \\
		proposed        & \textbf{0.317}         & \textbf{0.420}         & \textbf{0.459}         & \textbf{0.469}         \\
		\hline
    (A)  proposed,  no whitening                 & 0.310                  & 0.387                  & 0.409                  & 0.413                  \\
		(B)  proposed,  no ensembling              & 0.303                  & 0.404                  & 0.444                  & 0.453                  \\
		(C)  proposed,  observation-dependent     & 0.282                  & 0.396                  & 0.432                  & 0.441                  \\
		(D1) proposed,  no learnable speckles module & 0.265                  & 0.384                  & 0.426                  & 0.444                  \\
		(D2) proposed,  no learnable object module   & 0.312                  & 0.389                  & 0.411                  & 0.408                  \\
		\bottomrule
	\end{tabular}
	\caption{Model ablation study on multiple settings, see Sect. \ref{subsec:ablation}. As a comparison, the performance of the non-ablated \thisalgo and PACO algorithms are given on the first two lines. Reported results are the AUC averaged over the nine datasets \#1 to \#9 presented in Sect. \ref{subsec:detection_synthetic_sources}. 
  The best results are emphasized in bold fonts.}
	\label{tab:ablation_main_results}
\end{table*}

\begin{table}
	\centering
	\begin{tabular}{ccccc}
		\toprule
		                     & $T_{\textrm{test}}=16$ & $T_{\textrm{test}}=32$ & $T_{\textrm{test}}=48$ & $T_{\textrm{test}}=64$ \\
		\midrule
		$T_{\text{clip}}=16$ & 0.315                  & 0.406                  & 0.436                  & 0.445                  \\
		$T_{\text{clip}}=32$ & 0.318                  & 0.420                  & 0.454                  & 0.465                  \\
		$T_{\text{clip}}=48$ & \textbf{0.318}         & \textbf{0.421}         & \underline{0.458}      & \underline{0.468}      \\
		$T_{\text{clip}}=64$ & \underline{0.317}      & \underline{0.420}      & \textbf{0.459}         & \textbf{0.469}         \\
		\bottomrule
	\end{tabular}
	\caption{Model ablation study on setting (F): influence of the length of the clips $T_{\text{clip}}$ at training time versus of the number $T_{\text{test}}$ of frames in the test datasets. Reported results are the AUC averaged over the nine datasets \#1 to \#9 presented in Sect. \ref{subsec:detection_synthetic_sources}. 
  The best results are emphasized in bold fonts and the second best results are underlined.}
	\label{tab:ablation_clip}
\end{table}

\begin{figure}
	\centering
	\includegraphics[width=.475\textwidth]{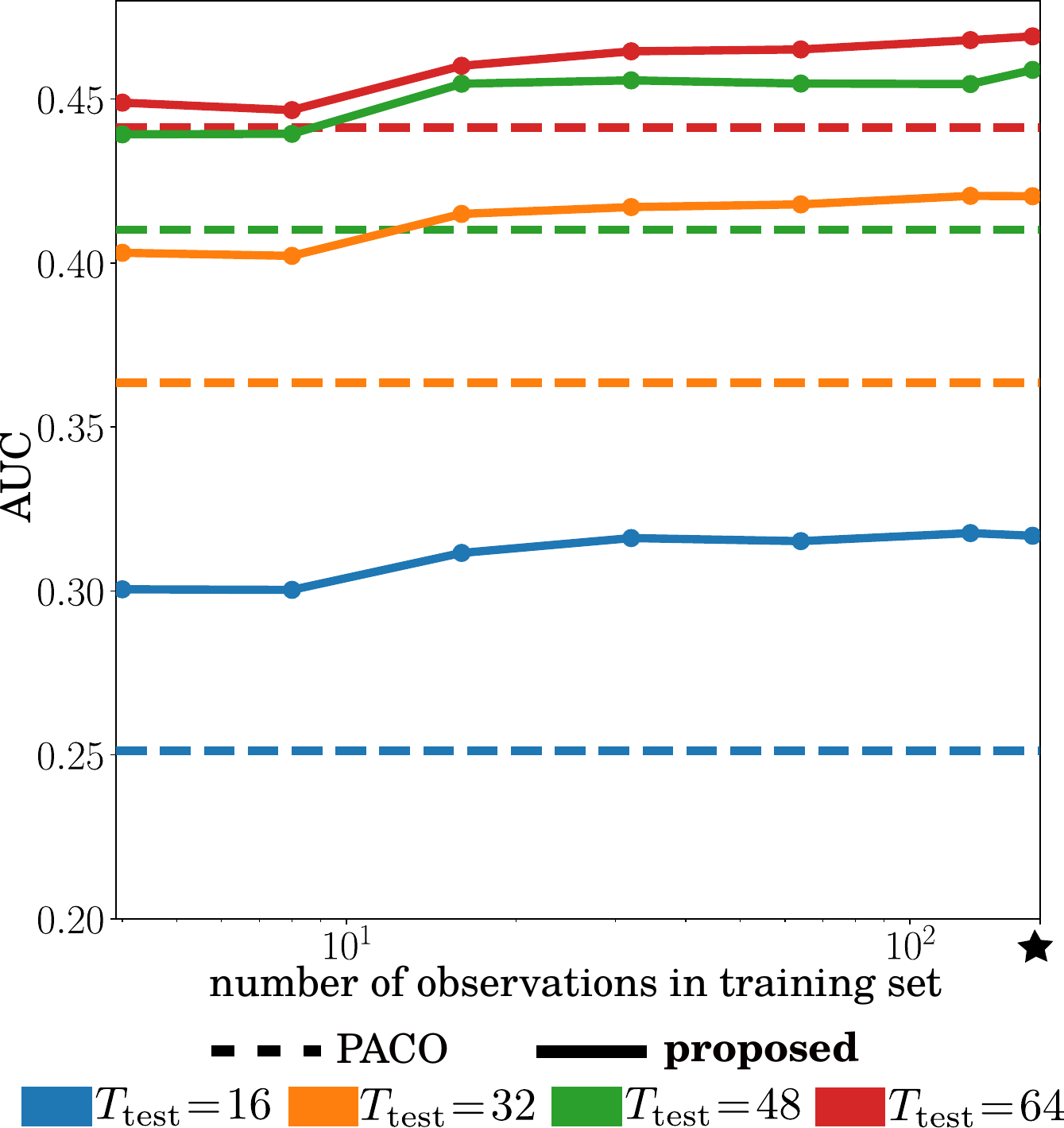}
	\caption{Model ablation study on setting (E): influence of the number of observations considered at training time for datasets having different number $T_{\text{test}}$ of frames at testing time. 
		As a comparison, the performances of the PACO algorithm are also given. 
Reported results are the AUC averaged over the nine datasets \#1 to \#9 considered in Sect. \ref{subsec:detection_synthetic_sources}.}
The proposed algorithm has been trained with all the observations of the training set (165).
	\label{fig:n_obs_auc}
\end{figure}

\begin{table*}
	\centering
	\begin{tabular}{ccccccc}
		\toprule
		algorithm & train spectral band & test spectral band & $T_{\textrm{test}}=16$ & $T_{\textrm{test}}=32$ & $T_{\textrm{test}}=48$ & $T_{\textrm{test}}=64$ \\
		\midrule
		PACO      & -                  & H2                 & 0.251                  & 0.363                  & 0.410                  & 0.441                  \\
		proposed  & H2-H3               & H2                 & 0.317                  & 0.420                  & 0.459                  & 0.469                  \\
		\midrule
		proposed (ablated) & H2                  & H2                 & 0.317                  & 0.422                  & 0.459                  & 0.469                  \\
		proposed (ablated) & H2-H3 \& K1-K2      & H2                 & 0.315                  & 0.419                  & 0.454                  & 0.463                  \\
		\bottomrule
	\end{tabular}
	\caption{Model ablation study on setting (G): influence of the spectral band(s) used for training. Evaluation is systematically done on the H2 band. As a comparison, the performance of the non-ablated \thisalgo and PACO model are given on the first two lines. Reported results are the AUC averaged over the nine datasets \#1 to \#9 presented in Sect. \ref{subsec:ablation}.}
\end{table*}

We present in this section an \gls{ablation_study} of \thisalgo (i.e., we downgrade it progressively) to evaluate the benefits of its main components.
We thus successively discuss the influence of modeling the covariances of the learned features (denoted ablation study A in the following), and the combination of multiple reconstructions by model ensembling (B).
More generally, this study also aims to determine which information contained in high-contrast observations is the most critical to model accurately the nuisance component.
For that purpose, we compare the proposed approach to an equivalent observation-dependent model (C).
We also investigate in studies (D1) and (D2) the benefits of each of the two learnable modules working on two complementary representations of the data.
Additionally, we assess the influence of the total number of training datasets (E) and the length of clips used to train our model (F).
Finally, we study the influence of the considered spectral bands in the training set (G).

Concerning point (A), the downgraded model is obtained by ignoring the spatial covariances in the whitening procedure embedded within the residual block of the speckles-aligned learning module, see Fig. \ref{fig:architecture}. 
In practice, we enforce $\widehat{\rho}=1$ and thus $\widehat{\M C} = \widehat{\M Q}$ in Eq. (\ref{eq:shrinkage_convex_combination}). 
The variances are still accounted for in the resulting covariance matrix $\widehat{\M C}$, thus acting as a classical normalization of the data, see Sect. \ref{subsubsec:speckles_aligned_stage}.
Concerning point (B), we compare a single prediction from a single model (i.e., $M=1$) to the final output of the proposed algorithm combining $M=10$ individual reconstructions, see Sect. \ref{subsucsec:model_ensembling}.
For ablation (C), we train the proposed architecture from a single dataset corresponding to the target dataset from which we also aim to detect unknown sources at testing time, as done by any observation-dependent model.
For studies (D1) and (D2), we compare the influence of the model architecture for a fixed amount of training data (the full database, as with an observation-independent model).
We consider two downgraded models, one comprising only the speckle-aligned module and the other comprising only the object-aligned module.
In order to reconstruct the target signal $\V x$ from the speckles-aligned block only, we add after the patch aggregation and feature derotation operations (see Fig.
\ref{fig:architecture}), a linear projection and a non-linearity in the form of a \gls{rectified_linear_unit} function with a learnable threshold.
Conversely, to estimate $\V x$ from the objects-aligned block only, we consider pre-processed observations that are whitened for the spatial covariances as input of the associated U-Net learnable module.
The resulting ablated model is quite similar to the deep PACO architecture, except that it is trained on more data than deep PACO, and that the input of the U-Net is formed by a single frame (after pixel-wise summation along the temporal dimension) instead of the full set of $T$ pre-processed frames as in deep PACO.
On the latter point, we also considered these variants in the deep PACO algorithm, and we experienced very similar results with each of them \citep{flasseur2024deep}, so that our ablation study is fair to compare the influence of the model architecture.
In (E), we study the impact of the number of observations used to train the model.
The observations are randomly selected from the full training database, regardless of the number $T$ of frames within each dataset.
In (F), we investigate the effect of the length of the clips $T_\text{clip}$ at training time on performances at inference time.
Finally, we study in (G) the influence of the spectral bands considered at training and testing times. 
Specifically, we investigate whether it is beneficial to train a model for each spectral band individually or to use a single model trained on all spectral bands.

Table \ref{tab:ablation_main_results} summarizes the main results of ablations (A-D). 
Scores are presented as in Sect. \ref{subsec:detection_synthetic_sources}, i.e. we report the AUC metric averaged over the same test datasets as in Sect. \ref{subsec:detection_synthetic_sources}. 
Concerning model ablation (B) related to the ensembling strategy, Appendix \ref{app:ablation_details} reports additional results in the form of AUC scores as a function of the number $M$ of ensembled models, but for different test datasets having varying numbers $T_{\text{test}}$ of temporal frames.
Table \ref{tab:ablation_clip} reports similar scores for ablation (F) on the relative influence of $T_\text{clip}$ and $T_\text{test}$. 
Figure \ref{fig:n_obs_auc} reports results obtained by varying the number of observations in the training set.

Based on these results, the main conclusions are as follows.
The whitening procedure applied to the learned features is beneficial to the proposed approach.
It also means that the associated whitening (of the data or of the associated features) through the covariances can not be fully replaced by a learnable module, because of the matrix inversion and factorization it implies (see Eq. (\ref{eq:whitening})) is difficult to implement through the linear operations (namely, convolutions) and non-linearity (namely, activation functions) of the trainable stages.
The model ensembling strategy applied to the output of the algorithm also improves the performance of the proposed approach, mainly by mitigating false alarms hallucinated by the network.
Additional results reported in Appendix \ref{app:ablation_details} illustrate that averaging results over $M=10$ different models is sufficient, as a plateau is reached in terms of AUC.
Ablation (C) clearly demonstrates that useful information can be learned from a training database of multiple observations.
Indeed, keeping the same architecture as the proposed one, but estimating its underlying parameters $\V \theta$ from a single dataset (corresponding to the target dataset itself) leads only to a slight gain with respect to PACO.
Ablation (D) complements these lessons, by showing that useful information can be extracted from both the speckles-aligned and object-aligned representation of the data, which justifies the choice of the proposed architecture with two learnable modules working on these two complementary views of the data.
Interestingly, suppressing the speckles-aligned module (thus corresponding to a very similar architecture to deep PACO) leads only to a slight gain with respect to PACO, for the range of angular separations considered in this paper.
Ablation (E) concerning the data volume at training time illustrates that several training datasets are needed to capture the variability of the speckles in the testing dataset (not seen at training time).
However, the number of datasets required to capture this diversity is limited, as a plateau in terms of AUC is reached when more than approximately 60 datasets are included in the training base.
Interestingly, even for smaller numbers of datasets in the training base, the proposed approach outperforms PACO.
Beyond that, when the training database contains more than 10 to 20 different datasets, the proposed approach trained on datasets of $T_{\text{train}}=32$ (respectively, $T_{\text{train}}=48$ frames) leads to better performance than PACO applied on the same target datasets but with 16 additional temporal frames.
Ablation (F) illustrates that the overall performances degrade when the number of frames $T_{\text{test}}$ of the test datasets is higher than the length of clips $T_{\text{clip}}$ used at training time.
We selected $T_\text{clip}=64$ as this leads to the best trade-off between performances and versatility, considering the distribution of the number of frames shown in Fig. \ref{fig:database_a}.
Finally, ablation study (G) on the spectral bands shows, as expected, that the best performance of \thisalgo is obtained when the training database contains datasets from the same spectral band. 
Including additional spectral bands (for instance, using both H and K bands when the model is tested solely on the H band) slightly degrades the detection performance. 
These results justify our design choice of using a band-dependent model.
However, including the H2-H3 dual bands during training, while applying the model only on the H2 band during testing, does not significantly alter the results compared to training solely on the H2 band. This can be understood as the relationship between the H2 and H3 bands can be approximated by a homothety. Consequently, training on the H2-H3 dual band does not introduce significantly more diversity than training solely on the H2 band.

\subsection{Sensitivity to observing conditions}
\label{subsec:obs_cond}

In this section, we evaluate the performance of our model with different observing conditions, both at training and test time.

The observations from the F150 archive were obtained under a diverse range of observing conditions (see Fig. \ref{fig:database}(d)), which affects the quality of the resultant data \citep{courtney2023empirical}.
For instance, the \textit{wind-driven halo} is a well-known artifact that can manifest when atmospheric turbulence conditions fluctuate more rapidly than the adaptive optics system can correct (\cite{cantalloube2020wind}).
Conversely, the \textit{low-wind effect} (LWE) emerges under conditions of relatively low wind speed. This phenomenon can be attributed to temperature inhomogeneities across the telescope's pupil (\cite{sauvage2015low, milli2018low}).
In addition to the observing conditions, other parameters affect the quality of a measurement such as the star magnitude (as it affects the quality of the adaptive optics correction; \cite{cantalloube2019peering}), but the latter are not considered in this section.


To quantify the quality of a dataset, we use the average \textit{astronomical seeing} and \textit{coherence time} for each observation from the atmospheric site monitor\footnote{\href{https://www.eso.org/sci/facilities/paranal/astroclimate/asm-instruments.html}{https://www.eso.org/sci/facilities/paranal/astroclimate/asm-instruments.html}} of the VLT, and
we categorize the observations into four groups representing different observing conditions.
We isolate the measurements affected by the low-wind effect as these datasets can be misclassified
 as good observations based on seeing and coherence time measurements only.
According to \cite{milli2018low}, LWE can be identified by the ambient wind speed: less than \unit{3}{\meter\per\second}
if the measurement occurred before November 19, 2017; less than \unit{1}{\meter\per\second} thereafter.
We have used the seeing and coherence time to divide the remaining observations into three categories of similar sizes, designated as \textit{bad}, \textit{average}, and \textit{good} in the following. 
The distribution of the resulting classes as a function of the seeing and coherence time is shown in Fig. \ref{fig:database}(d).

In order to compare the performances of our method in various observing conditions, we split both train and test sets into \textit{bad}, \textit{average}, and \textit{good} classes.
For each category, we trained an ensemble of models, with the same settings as described in Sect. \ref{sec:method}.
For evaluation, we pick test observations from each subset (one for \textit{bad}, two for \textit{average}, and two for \textit{good}).
Different from Sect. \ref{subsubsec:position_flux_sampling}, we inject sources in a fixed range of flux $[3 \times 10^{-6}, 3 \times 10^{-5}]$ (uniformly in log space).
This approach brings out the intrinsic difficulty (due to the varying observing conditions) associated with each dataset.
Sources are injected with multiple parallactic rotation amplitudes as in \ref{subsec:detection_synthetic_sources}, and results were averaged.
The flux and initial positions of sources are identical for all observations.

Table \ref{table:obs_cond} presents the results of this analysis, from which we draw several conclusions.
First, it is always better to train a unique model on all observations instead of training different models for different observing conditions.
Indeed, splitting the dataset per observing conditions reduces the number of observations available at training time, respectively 46, 43 and 44 observations for \textit{bad}, \textit{average} and \textit{good} conditions, 
versus 165 observations in total for the proposed model.
As such, splitting the training set by observing conditions reduces the number of observations available for each model.
This is also consistent with Fig. \ref{fig:n_obs_auc}, which shows that performances increase with the number of observations in the training set.

Second, the models trained under \textit{bad} and \textit{good} conditions exhibit poorer performance.
This result can be interpreted by considering that \textit{bad} observing conditions yield speckles with significantly different behavior, as detailed in \cite{cantalloube2019peering}.
In order to perform well in this setting, the model might learn some feature representation that does not transfer well to other conditions.
Conversely, the model trained only in \textit{good} conditions may lack diversity and robustness when applied on worse conditions.
The model trained on the \textit{average} subset presents the best balance among all models trained on subsets.

Finally, as previously noted, the performance gap between \thisalgo and PACO widens as the number of frames and the parallactic rotation rate (not represented in Table \ref{table:obs_cond}) decreases.

\begin{small}
\begin{table*}\centering
	\begin{tabular}{lccccccccccccc}\toprule
		                 & \multicolumn{3}{c}{$T_{\text{test}}=16$} & \multicolumn{3}{c}{$T_{\text{test}}=32$} & \multicolumn{3}{c}{$T_{\text{test}}=48$} & \multicolumn{3}{c}{$T_{\text{test}}=64$} & MEAN                                                                                                                                                                     \\
		\cmidrule(lr){2-4}\cmidrule(lr){5-7} \cmidrule(lr){8-10}\cmidrule(lr){11-13}
		Method (\textit{obs. cond.})        & \textit{bad}                        & \textit{average}                     & \textit{good}                       & \textit{bad}                        & \textit{average}             & \textit{good}               & \textit{bad}                & \textit{average}             & \textit{good}               & \textit{bad}                & \textit{average}             & \textit{good}               & \\
		\midrule
		Ours (\textit{bad})        & \underline{0.406} & 0.519             & \underline{0.594} & 0.491             & 0.551             & 0.650              & 0.532             & 0.561             & 0.653             & 0.554             & 0.559             & 0.666             & 0.561 \\
    Ours (\textit{average})     & \underline{0.406} & \textbf{0.526}    & \underline{0.594} & \underline{0.494} & \underline{0.559} & \underline{0.654}  & \underline{0.533} & \underline{0.572} & \underline{0.664} & \underline{0.556} & \underline{0.572} & \underline{0.679} & \underline{0.567} \\
		Ours (\textit{good})       & \underline{0.406} & 0.520             & 0.588             & 0.485             & 0.556             & 0.647              & 0.531             & 0.566             & 0.653             & 0.551             & 0.563             & 0.672             & 0.561\\
    Ours (all+LWE)  & \textbf{0.419}    & \underline{0.523} & \textbf{0.598}    & \textbf{0.500}    & \textbf{0.559}    & \textbf{0.655}     & \textbf{0.542}    & \textbf{0.572}    & \textbf{0.667}    & \textbf{0.563}    & \textbf{0.573}    & \textbf{0.680}    & \textbf{0.571} \\
		PACO              & 0.325             & 0.461             & 0.516             & 0.449             & 0.493             & 0.599              & 0.503             & 0.521             & 0.636             & 0.537             & 0.532             & 0.659             & 0.519\\
		\bottomrule
	\end{tabular}
	\caption{
		Ablation study on observing conditions:
		we conducted a comparative analysis to assess the performances of our method trained (lines) and evaluated (columns) on datasets recorded under different observing conditions.
		The specific subset used for training is indicated in parentheses for each instance of our method.
		We evaluated the AUC metric (see Sect. \ref{subsec:evaluation_metrics}) of each method across test observations from the \textit{bad}, \textit{average} and \textit{good} subsets. As a comparison, the performance of the non-ablated \thisalgo (trained on all considered datasets, regardless of the experienced observing conditions) and of PACO models are given on the first two lines. 
		Sources were injected with varied numbers $T_{\text{test}}$ of frames and parallactic rotation amplitudes $\Delta_\phi$. The reported results are averaged over the parallactic rotation amplitudes. 
	}
  \label{table:obs_cond}
\end{table*}
\end{small}

\subsection{Detection of known real sources}
\label{subsec:detection_real_sources}

\begin{figure*}
	\centering
  \includegraphics[width=0.95\textwidth]{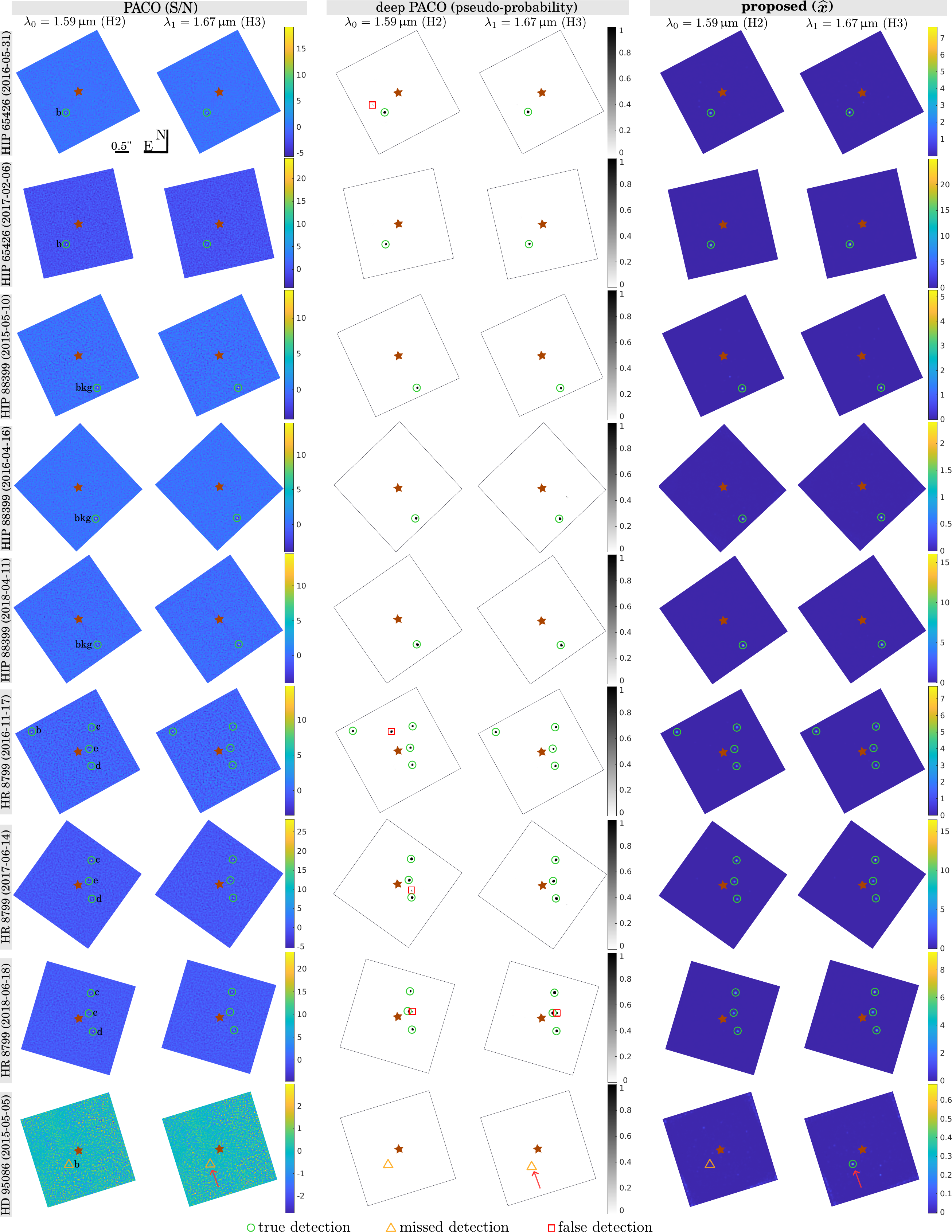}
	\caption{
		Detection maps obtained in the H2 and H3 spectral bands with PACO (1\textsuperscript{st} and 2\textsuperscript{nd} columns), deep PACO (3\textsuperscript{rd} and 4\textsuperscript{th} columns), and the proposed \thisalgo algorithm (5\textsuperscript{th} and 6\textsuperscript{th} columns).
		For each algorithm, the two spectral bands are processed separately.
		For each algorithm and dataset, the dynamic of the display is common for the two spectral channels: it is set between the minimum and maximum values of the two resulting detection maps.
    The colormap used for deep PACO is different to emphasize that its output is bounded within $[0, 1]$, as it represents a pseudo-probability.
    Known real sources, either exoplanets (\textit{b, c, d, e}) or faint background sources (\textit{bkg}) within the field of view are specified.
    We used a detection threshold of 4 for PACO and \thisalgo, and 0.5 for deep PACO.
    An interesting case, where only \thisalgo is able to detect a known real source is indicated with a red arrow.
		See \ref{subsec:datasets} for the observation parameters.
  }
	\label{fig:real_sources_detection_maps_fullfig_same_dyn}
\end{figure*}

In this section, we evaluate qualitatively the ability of the proposed approach to (re)detect known real sources, either exoplanets of emblematic systems or faint background sources in the field of view of imaged stars. 
For that purpose, we consider nine datasets of SPHERE-IRDIS (numbered \#10 to \#18 in Sect. \ref{subsec:datasets}). 
Figure \ref{fig:real_sources_detection_maps_fullfig_same_dyn} shows the detection maps ($\widehat{\V x}$) produced by the proposed approach, compared to the S/N map of PACO and the pseudo-probability
\footnote{We refer to the term \textit{pseudo-probability} as each pixel value of the detection map represents a score between 0 and 1 such that a high (respectively, a low) score values the presence (respectively, the absence) of a source centered at that location. 
However, this score can not be interpreted as a true probability of presence of a source as this would require a control of the uncertainties with dedicated methods, not implemented within deep PACO, see \cite{flasseur2024deep}.} map produced by deep PACO. 
Experiments are conducted separately on the two available spectral channels (H2 and H3). 
Overall, the proposed approach is the only one able to re-detect, at least in one of the two spectral channels, all the known sources (in the sense that there is no false alarm associated with a higher detection confidence than the real sources). 
Besides, translating the $\widehat{\V x}$ quantity produced by the proposed approach into a S/N of detection through the calibration procedure described in Sect. \ref{subsec:calibration} leads to a detection confidence better than the standard $5\sigma$ detection threshold for all sources excepted HD 95086 b (2015-05-05). 
However, in the latter example, HD 95086 b is detected in the H3 band at an equivalent S/N of 4.1 with the proposed approach while the source remains undetectable with PACO (S/N of 1.5) and this source is not detectable at all with deep PACO (but no false alarms are present in the field of view). 
The deep PACO algorithm discards most false alarms associated with low S/N with PACO as the method produces, by construction, a quasi-binary detection map. 
As a drawback of this peculiarity, the stability of the training and of the results at inference time strongly depends on the quality of the dataset. 
As an illustration, a bright detection blob (i.e., pseudo-probability close to 1) is observed on the detection map of HR 8799 (2016-11-17) with deep PACO. 
This detection is very likely a false alarm since it corresponds to an area impacted by outlier data in the science frames and, in the case of joint spectral processing (not shown here), this side detection blob disappears with deep PACO and is attenuated with PACO. 
In this respect, the proposed approach is more robust and resilient to the quality of the observations. 
As another example, the exoplanet HR 8799 e is detected with a very high S/N with PACO, with a ring around the main detection peak resulting from the cross-correlation of the first Airy lob of the off-axis PSF. 
As deep PACO is trained to detect blobs, part of this signature is interpreted as a second individual detection, while being part of the same source within the data. 
This effect is not observed on the reconstructed map produced by the proposed approach, which demonstrates again its improved robustness. 

Overall, the proposed approach significantly enhances detection sensitivity compared to the two reference methods, particularly in low parallactic rotation regimes where it achieves up to a tenfold increase in contrast in some cases.
It also demonstrates greater robustness, particularly compared to deep PACO, in handling outlier data and variability in observations.

%% file: conclusion.tex
\section{Discussion and conclusion} 
\label{sec:conclusion}

\noindent We presented a new post-processing algorithm for exoplanet detection from ADI observations. 
It leverages a database of multiple observations to build a deep non-linear model of nuisances (mainly quasi-static speckles) corrupting point-like object signals. 
By injecting simulated point-like sources into augmented observations, the model is trained to reconstruct a target quantity linked to the flux of the objects. 
Multiple reconstructions are combined by model-ensembling for improved robustness. 
The algorithm includes statistical modeling of spatial correlations at the scale of patches, inherited from the PACO algorithm. 
This modeling enhances both data contrast and stationarity, which have been shown to be crucial prerequisites for training our deep model given the noise statistics.
Estimation of local covariances and data whitening are incorporated into the learnable modules. 
Residual nuisances not captured by the statistical model (e.g., large stellar leakages) are handled by two trainable stages, which work on complementary data representations with either quasi-static speckles or off-axis sources aligned along the temporal dimension.

We evaluated the performances of the proposed approach using standard metrics on several observations from the VLT/SPHERE-IRDIS instrument. 
Our experiments showed that \thisalgo achieves a better trade-off between precision and recall than the state-of-the-art PACO algorithm. 
The gain is more pronounced when the diversity induced by ADI is limited, for example in the case of a small total parallactic rotation or few exposures. 
In such cases, observation-dependent post-processing algorithms suffer from self-subtraction, leading to a bias in source flux estimation and limiting detection performance. 
The proposed approach partly mitigates this limitation by learning robust features across multiple observations. 
In terms of achievable contrast, we observed a gain by a factor eight to ten (between 0.15" and 1.5") brought by the proposed approach with respect to PACO for datasets having a parallactic rotation of a few degrees. 
For moderate parallactic rotation, this gain typically manifests itself only below 0.35". 
For parallactic rotation typically higher than 30 degrees, the proposed approach and PACO lead to comparable detection sensitivity when the observing conditions are good. 
Nevertheless, the proposed approach is more robust against false alarms than PACO and deep PACO, especially when the number of frames is limited (because, in this setting, the covariances are shrunk towards zero by the two comparative algorithms) or when the observing conditions are bad. 
These advantages could prove valuable for scrutinizing the inner environments of nearby solar-type stars using upcoming thirty-meter class telescopes (e.g., ELT, GMT, TMT). 
In these scenarios, the angular diversity of observations may be constrained by small angular separations of the sought objects, and total exposure time could be limited due to highly competitive access to observational resources.

Through a model ablation study, we have identified the principal methodological components that contribute most significantly to improving the precision-recall trade-off in our approach.
Our results underscore the importance of leveraging information from an archive of multiple observations to construct a refined model of nuisances. This approach allows us to effectively address the variability in observational conditions.
Furthermore, we validate the efficacy of employing two distinct learnable modules that operate on different data representations—temporally co-aligned speckles and off-axis sources. These modules capture information across varying spatial scales, thereby enhancing overall detection performance.
The robustness and stability of our proposed algorithm against data heterogeneity, arising from diverse observing conditions, are ensured through:
(i) a sophisticated loss function that adapts to the challenge of source detection considering factors such as flux, angular separation, and local statistical characteristics of nuisances;
(ii) ensemble techniques that aggregate multiple network predictions to mitigate potential false alarms;
(iii) modeling covariances of learned features, which is crucial as disregarding these correlations significantly impacts detection performance.
In summary, our study underscores the significance of leveraging diverse observational data, employing multiple learning modules, and rigorously modeling feature covariances to achieve robust and effective source detection in complex observational settings.

As methodological improvements, we are currently investigating the main limitation of the proposed approach, namely the lack of built-in access to a fully unbiased estimation of the flux distribution and of the associated uncertainties. 
The proposed algorithm could also be extended to the joint processing of multi-spectral datasets such as the ones provided by the VLT/SPHERE-IFS instrument using the angular plus spectral differential imaging technique. 
Beyond those aspects, the observation-independent model of the nuisance is general and can thus be employed to solve other tasks like the reconstruction of the spatial flux distribution of circumstellar environments.

%% file: appendix.tex





\section{Glossary}
\label{app:glossary}

The following glossary provides definitions of key terms and technical concepts from the machine learning literature referenced throughout the paper.
\vspace{-25pt}
\glsnogroupskiptrue
\printglossary[nonumberlist,title=]


\section{Detailed ROCs results}
\label{app:detailed_rocs_results}

\begin{figure*}
	\centering
	\includegraphics[width=0.9\textwidth]{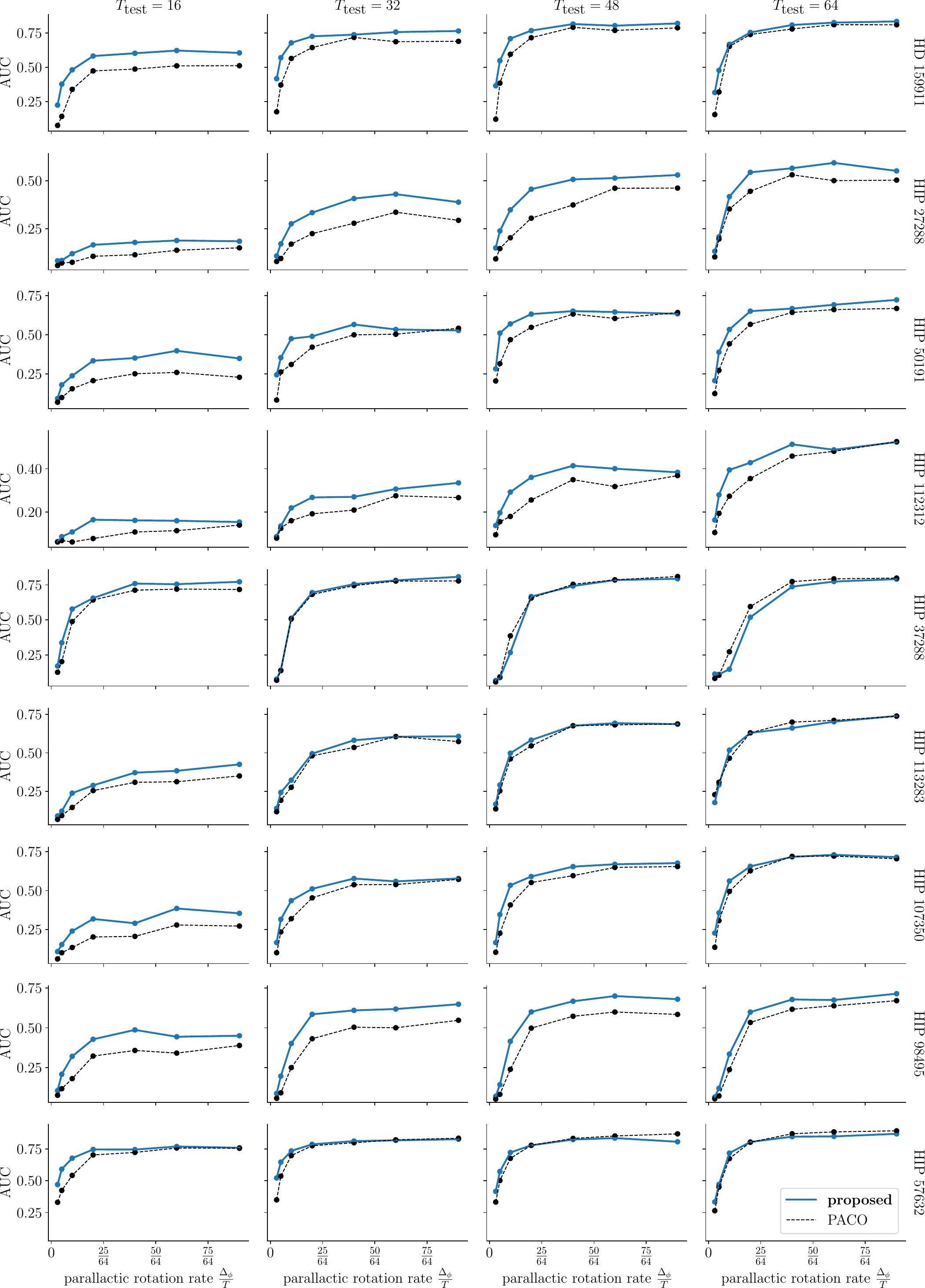}
  \caption{AUC obtained after injection of synthetic sources, for each of the nine observations \#1 to \#9 (from top to bottom) described in Table \ref{table:observations}. Different numbers $T_{\text{test}}$ of temporal frames are considered at testing time, see columns.}
	\label{fig:auc_fdr_tpr_detailed_fullfig}
\end{figure*}

\begin{figure*}
	\centering
	\includegraphics[width=0.9\textwidth]{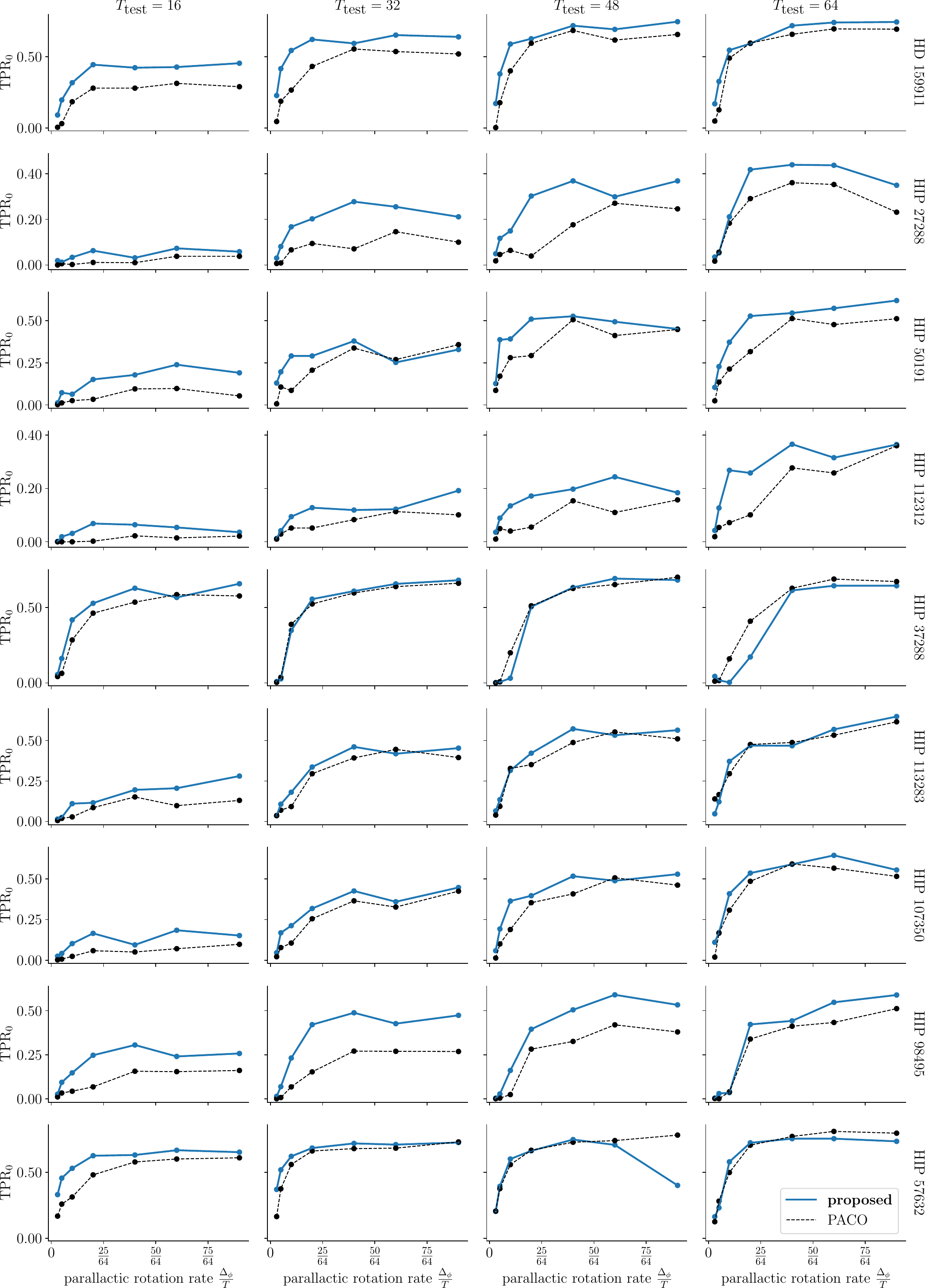}
  \caption{$\text{TPR}_0$ obtained after injection of synthetic sources, for each of the nine observations \#1 to \#9 (from top to bottom) described in Table \ref{table:observations}. Different numbers $T_{\text{test}}$ of temporal frames are considered at testing time, see columns.}
	\label{fig:tpr_max_detailed_fullfig}
\end{figure*}

In this appendix, we give additional results complementing Fig. \ref{fig:avg_auc_fdr_tpr_main_fullfig} with the detailed AUC results obtained on the nine SPHERE-IRDIS datasets \#1 to \#9 that were considered in Sect. \ref{subsec:detection_synthetic_sources}.
Figure \ref{fig:auc_fdr_tpr_detailed_fullfig} represents the resulting detection AUC (TPR versus FDR) as a function of the total amount $\Delta_{\phi}$ of parallactic rotation. 
As in Fig. \ref{fig:avg_auc_fdr_tpr_main_fullfig}, we build the experiments by keeping $T_{\text{test}} \in \llbracket 16, 32, 48, 64 \rrbracket$ temporal frames from the original datasets in order to study the influence of both the angular diversity and of the amount of available data on the performance of observation-dependent (PACO) and observation-independent (\thisalgo) models. 

 



\section{Additional results on ablation analysis}
\label{app:ablation_details}


\begin{figure}
	\centering
	\includegraphics[width=.475\textwidth]{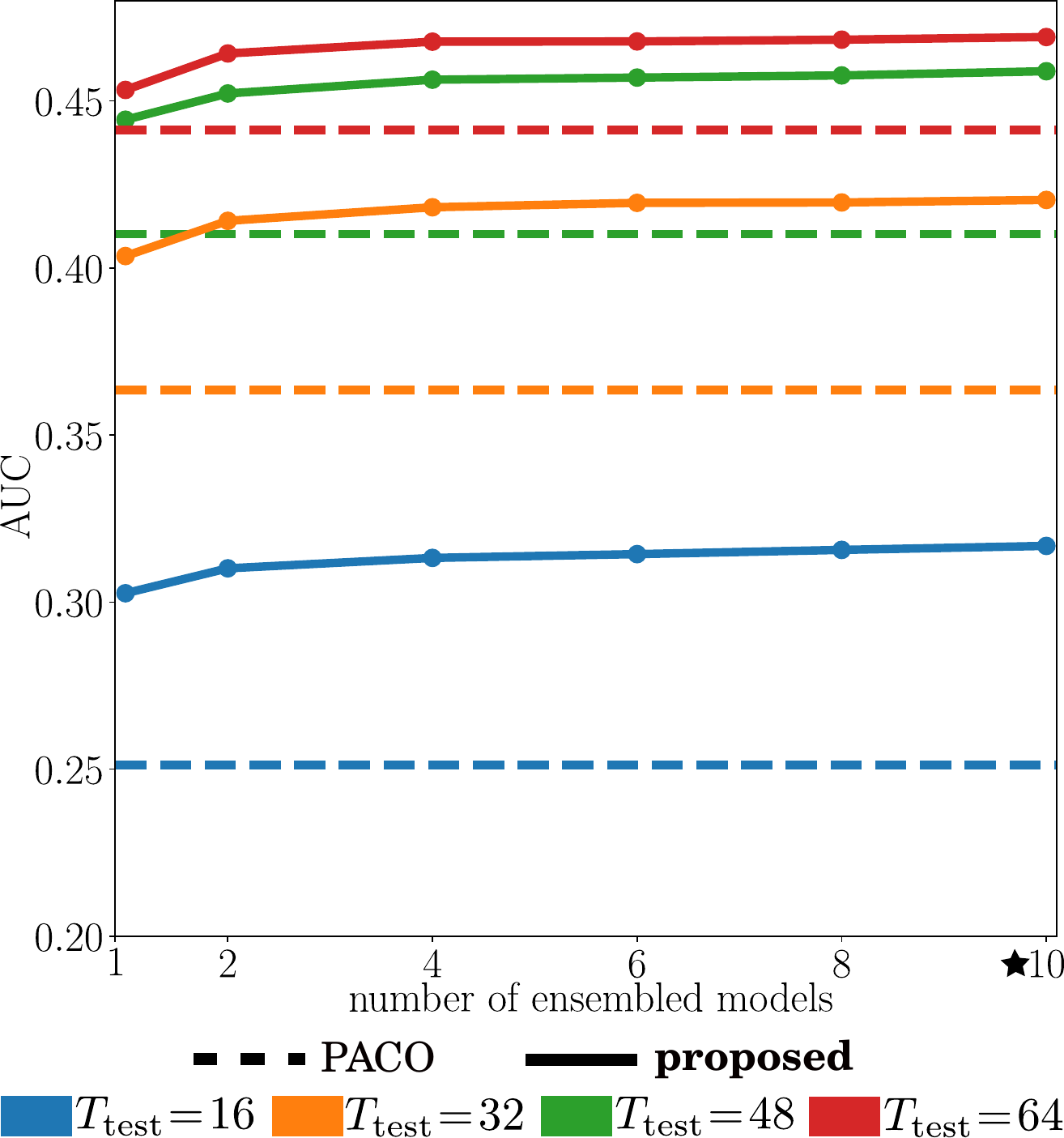}
  \caption{AUC versus number $M$ of ensembled models. The results are averaged over the nine observations \#1 to \#9 described in Table \ref{table:observations}, with seven different parallactic rotation rates, as in Sect.\ref{subsec:detection_synthetic_sources}.}
  \label{fig:ensembling}
\end{figure}

Figure \ref{fig:ensembling}, shows that the detection performance of the proposed approach increases with the number of ensembled models, until a plateau is reached. In all of our experiments, we used an ensemble of $M=10$ models, see Sect. \ref{subsec:ablation}.

\newpage